\documentclass[11pt,a4paper]{article}
\usepackage{amsmath,amssymb,amsthm}
\usepackage{jheppub}
\usepackage{graphicx}
\usepackage{hyperref}
\usepackage{xcolor}
\usepackage{booktabs}
\usepackage{siunitx}
\usepackage{subcaption}
\usepackage{float}
\usepackage[toc,page]{appendix}
\usepackage{tikz}
\usepackage{multicol}

\newcommand{\Mpl}{M_{\rm pl}}

\newcommand{\dd}[1]{\mathrm{d}#1}

\newcommand{\Seff}{S_{\rm eff}}
\newcommand{\Sgrav}{S_{\rm grav}}
\newcommand{\Sfl}{S_{\rm fl}}
\newcommand{\Spp}{S_{\rm pp}}
\newcommand{\Sint}{S_{\rm int}}

\newcommand{\kperp}{k_\perp}

\title{\boldmath Effective Field Theory of Gravity in Relativistic Media}

\author[a]{Beka Modrekiladze}

\affiliation[a]{Deutsches Elektronen-Synchrotron DESY,\\ Notkestra{\ss}e 85, 22607 Hamburg, Germany}

\emailAdd{beka.modrekiladze@desy.de}

\abstract{
We develop an effective field theory of gravity in relativistic media.
Integrating out the medium leaves vacuum gravity with updated Feynman rules:
a graviton propagator dressed by the stress-energy two-point function of the
environment, medium-induced bulk graviton vertices from higher correlators,
and generalized worldline couplings encoded in matched Wilson coefficients.
The diagram topologies are unchanged from vacuum, so different
media (perfect fluids, collisionless matter, coherent scalar fields such as wave dark
matter) are not different theories but different correlators
inserted into the same diagrams.

We derive the in-medium rules for a relativistic fluid and obtain the full
1PN Einstein--Infeld--Hoffmann potential, the 1PN Stokes drag, the gravitational self-energy and a generalized Christodoulou memory whose new tensor structure records the
orientation of the medium's anisotropy,
turning the permanent strain into an
\emph{astrophysical weathervane}. The same rules activate phenomena forbidden in vacuum: the sound pole converts the symmetry-protected, non-running black-hole Love number into a resonant,
running one, and 
graviton splitting $h\to hh$ is open into the longitudinal branch of the
dressed propagator, and gives the transverse graviton a width,
$\Gamma=G_N\omega^3(1-c_s^2)^2/120c_s^3$, a \emph{gravitational opacity} set by
the sound speed. The same vertex makes a graviton suffer dynamical
friction. 

We assess observational prospects, from dephasing and tidal
resonances within reach of the Einstein Telescope and LISA to
proof-of-principle memory and opacity signatures.
}

\begin{document}
\maketitle

\section{Introduction}
\label{sec:intro}

As gravitational-wave sensitivity improves with the Einstein Telescope~\cite{Punturo:2010zz,Maggiore:2019uih} and
LISA~\cite{LISA:2017pwj}, the astrophysical environments in
which compact binaries evolve---accretion disks, dark matter
spikes~\cite{Gondolo:1999ef,Merritt:2002vj}, bosonic condensates---will
leave increasingly visible imprints on the waveform. Dedicated programs
aim to extract these
signatures~\cite{Kavanagh:2020cfn,Coogan:2021uqv,Cole:2022yzw,Boskovic:2025ixx}.
Unmodeled, such corrections can mimic beyond-standard-model physics or
obscure genuine new physics~\cite{Gupta:2024gun}.

A systematic treatment of compact binaries in vacuum is provided by the
worldline EFT of Goldberger and
Rothstein~\cite{Goldberger:2004jt,Goldberger:2005cd}, which organizes the
post-Newtonian expansion diagrammatically and makes power counting
manifest. A program to promote this framework to compact objects embedded
in media was introduced in~\cite{Modrekiladze:2026kwt}. Two of its pieces are in place: the worldline theory of compact objects in relativistic
viscous fluids, with its matched one-body Wilson coefficients, was derived
in~\cite{Modrekiladze:2024htc}, and the in-medium graviton two-point
function, together with the self-force it generates, was obtained
in~\cite{Modrekiladze:2026twz}. Here we complete the program by writing
the full set of in-medium Feynman rules, including the medium-induced
three-graviton vertex, so that diagrams of the vacuum theory can be automatically evaluated in any medium.

The observation of this paper is that environmental physics can be
organized at the level of the Feynman rules. Once the medium is integrated
out, the only remaining bulk dynamical field is the graviton, and the
diagram topologies are the same as in vacuum. The environment enters
through a set of updated rules: a graviton propagator dressed by the
stress-energy two-point function of the medium, bulk graviton vertices
induced by higher stress-energy correlators, and generalized worldline
couplings encoded in matched Wilson coefficients. The medium correlators
and one-body matching data are \emph{computed once} and inserted into otherwise
standard gravitational diagrams: gravity in a medium is vacuum gravity
with replaced propagators and vertices.

We illustrate the replacement-rule logic through two classes of examples. 1) Diagrams that are
nonzero already in vacuum obtain medium corrections: the full 1PN
Einstein--Infeld--Hoffmann potential in a relativistic fluid, the gravitational self-energy and in-medium corrections to gravitational-wave memory. 2) Diagrams forced to
vanish in vacuum are activated by the medium: black-hole Love numbers and
graviton splitting, $h\to hh$, into the
longitudinal branch of the dressed propagator, which leads to
a gravitational opacity. The full causal self-force and the running of the drag coefficient are treated in the companion paper~\cite{Modrekiladze:2026twz}.

Sec.~\ref{sec:setup} builds the effective theory and its in-medium Feynman
rules, catalogues the media---perfect fluid, collisionless matter, coherent
scalar fields---unified by a single screening response, benchmarks against
the known Jeans dispersions (with wave-dark-matter observables in
App.~\ref{app:benchmarks}),
and identifies the induced renormalization-group flow. Secs.~\ref{sec:applications_nonzero} and
\ref{sec:applications_zero} treat topologies with nonzero and vanishing
vacuum limits, Sec.~\ref{sec:observations} assesses observational
prospects, and Sec.~\ref{sec:conclusions} concludes.

We use units $c=\hbar=1$, the mostly-minus metric convention, and
\begin{equation}
    \kappa \equiv \frac{2}{M_{\rm pl}},
    \qquad
    \kappa^2 = 32\pi G_N .
\end{equation}

\section{Effective Theory}
\label{sec:setup}

Our construction proceeds in two steps. We begin with compact objects
described by a worldline EFT coupled both to gravity and to a relativistic
medium described by its own bulk EFT. We then integrate out the medium.
The resulting theory contains only gravitons and worldlines, but with
propagators and vertices dressed by the response of the environment.

\subsection{Degrees of Freedom and Microscopic Action}

The long-distance degrees of freedom are the graviton $h_{\mu\nu}$, the
medium fields $\varphi$, and worldlines $x_i^\mu(\tau_i)$ describing the
compact objects. The observables studied below lie in a weak-gravity,
weak-background regime with the standard separation of scales: potential
gravitons ($k^0\sim v/r$, $|\mathbf k|\sim 1/r$) mediate post-Newtonian
interactions, while radiation gravitons ($k^0\sim|\mathbf k|\sim v/r$)
carry energy to infinity and enter observables such as memory and graviton
splitting. The expansion is controlled by
\begin{equation}
    \epsilon_{\rm grav} \equiv \frac{G_N M}{r} \ll 1,
    \qquad
    \epsilon_{\rm env} \equiv G_N \rho_E L^2 \ll 1,
\end{equation}
where $M$ is the compact-object mass, $\rho_E$ the ambient energy density
(the enthalpy $\epsilon+p$ for a relativistic fluid), and $L$ the
macroscopic scale of the background profile. A medium insertion
on a graviton line scales as $G_N\rho_E\ell^2$, with $\ell\sim r$
(potential region) or $\ell\sim r/v$ (radiation region), and remains
perturbative even where we package these corrections into exact dressed
building blocks. For a stable homogeneous background, $L$ cannot exceed
the Jeans scale, $\lambda_J\sim c_s/\sqrt{4\pi G_N(\epsilon+p)}$ for a
relativistic fluid. Finite-size effects are suppressed by powers of the
body size $R$ over the shortest macroscopic scale probed by the
worldline---in vacuum the familiar hierarchy behind $(R/r)^5$ tidal
effects---though in a medium relative boosts Lorentz-contract the
background scale seen by the body and can parametrically enhance them, so
that the effective finite-size parameter is $\epsilon_0\equiv\gamma
R/\alpha$, with $\alpha$ the environmental IR scale (for a background
density gradient, $\alpha\sim\rho_E/|\partial\rho_E|$).\footnote{We match
the notation of the companion paper~\cite{Modrekiladze:2026twz}, whose expansion
parameters are $\epsilon_0=\gamma R/\alpha$, $\epsilon_1=Gm\gamma/\alpha$,
and $\epsilon_2=G\rho_E L^2$: our $\epsilon_{\rm env}$ coincides with
$\epsilon_2$, while the post-Newtonian parameter $\epsilon_{\rm grav}$,
built on the orbital separation $r$, has no counterpart there and should
not be conflated with $\epsilon_1$.} We
work at the order where the leading Archimedean coefficients $F$ and $F'$
suffice.

\paragraph{Microscopic action.}
Before integrating out the medium, the long-distance theory is defined by
\begin{equation}
    S_{\rm tot}[x,h,\varphi]
    = \Sgrav[h] + \Sfl[\varphi] + \Spp[x,h] + \Sint[h,\varphi].
\end{equation}

\paragraph{Gravitational sector.}
We expand about flat space and fix de~Donder gauge. At quadratic order,
the Einstein--Hilbert action yields the vacuum graviton propagator
$D_{\mu\nu;\alpha\beta}$; higher terms generate the standard bulk graviton
self-interactions, beginning with the cubic vertex $V^{(3)}_{\rm EH}$.
In the post-Newtonian region, potential gravitons become instantaneous to
leading order, while radiation gravitons retain their full on-shell
kinematics.

\paragraph{Point-particle sector.}
Following~\cite{Modrekiladze:2024htc}, the worldline action of a compact
body moving through a relativistic medium is
\begin{equation}
    \Spp[x,h]
    = \int \dd{\tau}\,\sqrt{\dot x^2}\;
      F\!\left(\hat\gamma\right),
    \qquad
    \hat\gamma \equiv \frac{\dot x \cdot U}{\sqrt{\dot x^2}},
    \label{eq:Spp}
\end{equation}
where $U^\mu$ is the medium four-velocity evaluated on the worldline.
The function $F(\hat\gamma)$ packages the coupling of the compact object
to its environment. Expanding about vanishing relative velocity,
$\hat\gamma = 1$, its first Taylor coefficients are Wilson coefficients
matched in the single-body problem~\cite{Modrekiladze:2024htc}:
\begin{equation}
    F(1) = -M + \rho V_p,
    \qquad
    F'(1) = +\tfrac{3}{2}\rho V_p,
    \label{eq:F1}
\end{equation}
where $M$ is the gravitational mass, $\rho$ the local medium density, and
$V_p$ the effective volume of the body; in the vacuum limit
$F(1)\to -M$ and $F'(1)\to 0$. Where unambiguous we abbreviate
$F\equiv F(1)$ and $F'\equiv F'(1)$.

\paragraph{Medium--graviton interaction.}

The medium action depends on the metric, so gravity couples to the
medium at every order in $h_{\mu\nu}$,
\begin{align}
    \Sint[h,\varphi]
    &= -\frac{\kappa}{2}\int \dd{^4x}\,
      h_{\mu\nu}(x)\,\delta T^{\mu\nu}[\varphi](x)
    \notag\\
    &\quad
    + \frac{\kappa^2}{2}\int \dd{^4x}\dd{^4x'}\,
      h_{\mu\nu}(x)\,\mathcal S^{\mu\nu\alpha\beta}(x,x')\,h_{\alpha\beta}(x')
    + \mathcal O(h^3),
    \label{eq:Sint}
\end{align}

The linear term is the universal coupling to the stress tensor, with
$\delta T^{\mu\nu}$ the fluctuation about the chosen background. The
quadratic term is the seagull, and Eq.~\eqref{eq:seagull_def} gives
$\mathcal S^{\mu\nu\alpha\beta}$ explicitly.

\subsection{Integrating Out the Medium: The In-Medium Feynman Rules}
\label{sec:medium}

We now integrate out the bulk medium fields at fixed $h_{\mu\nu}$;
temporarily suppressing the worldline sector, the medium defines an
effective graviton action by
\begin{equation}
    e^{i\Seff[h]}
    \equiv
    \int \mathcal{D}\varphi\;
    \exp\!\big[\,
        i\Sfl[\varphi] {}+ i\Sint[h,\varphi]
    \,\big].
    \label{eq:Seff_def}
\end{equation}
Writing expectation
values in the medium background as $\langle\cdots\rangle_\varphi$

and expanding in powers of $h_{\mu\nu}$,
\begin{align}
    &\left\langle
        \exp\!\big[\,
            i\Sint[h,\varphi]
        \,\big]
       \right\rangle_{\!\varphi}
    \notag\\
    &\quad = 1
    - \frac{i\kappa}{2}\int \dd{^4x}\,
      h_{\mu\nu}(x)\,
      \langle \delta T^{\mu\nu}(x)\rangle_\varphi
    \notag\\
    &\quad
    + \frac{i\kappa^2}{2}
      \int \dd{^4x}\dd{^4x'}\,
      h_{\mu\nu}(x)\,
      \langle \mathcal S^{\mu\nu\alpha\beta}(x,x')\rangle_\varphi\,
      h_{\alpha\beta}(x')
    \notag\\
    &\quad
    - \frac{1}{2}\!\left(\frac{\kappa}{2}\right)^{\!2}
      \int \dd{^4x}\dd{^4x'}\,
      h_{\mu\nu}(x)\,h_{\alpha\beta}(x')\,
      \langle
        \delta T^{\mu\nu}(x)\,
        \delta T^{\alpha\beta}(x')
      \rangle_\varphi
    \notag\\
    &\quad
    + \frac{i}{3!}\!\left(\frac{\kappa}{2}\right)^{\!3}
      \int \dd{^4x}\dd{^4x'}\dd{^4x''}\,
      h_{\mu\nu}(x)\,h_{\alpha\beta}(x')\,h_{\rho\sigma}(x'')\,
      \langle
        \delta T^{\mu\nu}(x)\,
        \delta T^{\alpha\beta}(x')\,
        \delta T^{\rho\sigma}(x'')
      \rangle_\varphi
    \notag\\
    &\quad + \mathcal{O}(h^4).
    \label{eq:exp_expand}
\end{align}
The linear term is the tadpole, the quadratic term will dress the graviton
propagator, and the cubic term will induce a new bulk graviton vertex.

The full medium stress tensor decomposes as
$T^{\mu\nu}=\bar T^{\mu\nu}+\delta T^{\mu\nu}$, where the background piece
$\bar T^{\mu\nu}$ sources the equilibrium geometry; in the weak-background
regime $\epsilon_{\rm env}\ll 1$ that geometry is flat space, and the
coupling in Eq.~\eqref{eq:Sint} is to the genuine fluctuation
$\delta T^{\mu\nu}$. For a homogeneous equilibrium background
$\langle\delta T^{\mu\nu}\rangle_\varphi=0$, so the linear term
in~\eqref{eq:exp_expand} drops out. Taking the logarithm
re-exponentiates the connected pieces, $\Seff[h]=\Seff^{(2)}[h]
+\Seff^{(3)}[h]+\mathcal O(h^4)$, with the quadratic and cubic terms
given by the connected two- and three-point functions together with
the contact terms of the same order.

\paragraph{Quadratic order: the graviton polarization tensor.}

At fixed medium fields the metric dependence of $\Sfl$ expands as
\begin{align}
    \Sfl[\varphi,\eta+\kappa h]
    &= \Sfl[\varphi,\eta]
    -\frac{\kappa}{2}\int \dd{^4x}\,h_{\mu\nu}\,T^{\mu\nu}
    \notag\\
    &\quad
    +\frac{\kappa^2}{2}\int \dd{^4x}\,\dd{^4x'}\,
      h_{\mu\nu}(x)\,\mathcal S^{\mu\nu\alpha\beta}(x,x')\,h_{\alpha\beta}(x')
    +\mathcal O(h^3),
    \notag\\
    \mathcal S^{\mu\nu\alpha\beta}
    &\equiv\frac{\delta^2\Sfl}{\delta g_{\mu\nu}\,\delta g_{\alpha\beta}}
      \bigg|_{g=\eta}.
    \label{eq:seagull_def}
\end{align}
The quadratic term of $\Seff$ then has two pieces, the connected two-point
function of $\delta T$ and the expectation value of the seagull
$\mathcal S$:

\begin{equation}
    \Seff^{(2)}[h]
    = \frac{1}{2}\!\left(\frac{\kappa}{2}\right)^{\!2}
      \int \dd{^4x}\dd{^4x'}\,
      h_{\mu\nu}(x)\,
      \Pi^{\mu\nu\alpha\beta}(x,x')\,
      h_{\alpha\beta}(x'),
    \label{eq:Seff_quad}
\end{equation}
where
\begin{equation}
    \boxed{\;
    \Pi^{\mu\nu\alpha\beta}(x,x')
    \equiv
    i\,
    \big\langle
      \delta T^{\mu\nu}(x)\,
      \delta T^{\alpha\beta}(x')
    \big\rangle^{\rm conn}_{\varphi}
    \;+\;4\,\big\langle\mathcal S^{\mu\nu\alpha\beta}(x,x')\big\rangle_\varphi
    \,.\;
    }
    \label{eq:Pi_def}
\end{equation}
$\Pi$ satisfies the in-medium Ward identity
\begin{equation}
    k_\mu\,\Pi^{\mu\lambda\alpha\beta}(k)
    = \eta^{\lambda\alpha}k_\mu\bar T^{\mu\beta}
    + \eta^{\lambda\beta}k_\mu\bar T^{\mu\alpha}
    - k^\lambda\,\bar T^{\alpha\beta},
    \label{eq:ward}
\end{equation}
which follows because a coordinate change acting on a background with
$\bar T^{\mu\nu}\neq0$ varies the tadpole at $\mathcal O(\xi h)$. The
seagull is the gravitational analogue of the diamagnetic term in the photon
self-energy of a plasma. For the
perfect fluid at tree level it is local, momentum independent, and fixed by
the equilibrium data ($w\equiv\epsilon+p$,
$\Delta^{\mu\nu}=\eta^{\mu\nu}-U^\mu U^\nu$):
\begin{align}
    4\langle\mathcal S^{\mu\nu\alpha\beta}\rangle
    =\Big[&\epsilon\,U^\mu U^\nu U^\alpha U^\beta
    + p\,\big(U^\mu U^\nu\Delta^{\alpha\beta}+\Delta^{\mu\nu}U^\alpha U^\beta\big)
    + (p-c_s^2 w)\,\Delta^{\mu\nu}\Delta^{\alpha\beta}
    \notag\\
    &- p\,\big(\Delta^{\mu\alpha}\Delta^{\nu\beta}+\Delta^{\mu\beta}\Delta^{\nu\alpha}\big)
    - p\,\big(U^\mu U^\alpha\Delta^{\nu\beta}+3\ \text{perms}\big)
    \Big]\,\delta^{(4)}(x-x')\,,
    \label{eq:seagull_fluid}
\end{align}
which reduces for $p\to0$ to the sum over dust particles of the worldline
seagull of Fig.~\ref{fig:EIH_diagrams}(e). Its transverse-traceless block is
$-2p$ per unit polarization, a $\Lambda$-type contact term cancelled by the
curvature of the background it accompanies. The right-hand side of
\eqref{eq:ward} and the seagull are real, so the anti-Hermitian
(dissipative) part of $\Pi$ is transverse and is carried by the correlator
alone.
Added to the vacuum quadratic graviton action, $\Pi$ shifts the inverse
propagator, $G^{-1}=D^{-1}-i(\kappa/2)^2\,\Pi$; the resulting dressed
propagator is Eq.~\eqref{eq:Gfull} below. The medium thus dresses the
graviton through a universal object determined 
by the stress-energy two-point function of the environment together with
its equilibrium stress tensor.

\paragraph{Cubic order: the stress-tensor three-point kernel.}
At the next order, the connected three-point function generates a new bulk
graviton interaction,
\begin{equation}
    \Seff^{(3)}[h]
    =
    \frac{1}{3!}\!\left(\frac{\kappa}{2}\right)^{\!3}
    \int \dd{^4x}\dd{^4x'}\dd{^4x''}\,
    h_{\mu\nu}(x)\,
    h_{\alpha\beta}(x')\,
    h_{\rho\sigma}(x'')\,
    \Gamma^{\mu\nu,\alpha\beta,\rho\sigma}(x,x',x''),
    \label{eq:Seff_cubic}
\end{equation}
where $\Gamma^{\mu\nu,\alpha\beta,\rho\sigma}(x,x',x'')\equiv
\langle\delta T^{\mu\nu}(x)\,\delta T^{\alpha\beta}(x')\,
\delta T^{\rho\sigma}(x'')\rangle^{\rm conn}_{\varphi}$
$+$ the cubic seagull completions $3\langle\delta T\,\delta\mathcal
S\rangle+\delta^3\Sfl/\delta g^3$, mirroring Eq.~\eqref{eq:Pi_def}; without
them the cubic Ward identity fails exactly as the quadratic one does.
In momentum space this induces a correction $V^{(3)}_\Pi$ to the vacuum
three-graviton vertex, made explicit below.

\paragraph{The rules at a glance.}
\label{sec:feynman}
Once the medium is integrated out, every calculation proceeds exactly as
in vacuum: the diagrams are the vacuum diagrams, and only three building
blocks change,
\begin{equation}
    \boxed{\;
    \underbrace{-M\;\longrightarrow\;F(\hat\gamma)}_{\text{worldline vertices}}\,,
    \qquad
    \underbrace{D\;\longrightarrow\;G}_{\text{internal graviton lines}}\,,
    \qquad
    \underbrace{V^{(3)}_{\rm EH}\;\longrightarrow\;V^{(3)}_{\rm EH}+V^{(3)}_\Pi}_{\text{bulk cubic vertex}}\,.\;}
    \label{eq:dictionary}
\end{equation}
The recipe is: draw the vacuum diagrams for the observable at hand, apply
the three replacements, and add nothing else. The first rule controls the
two-body potentials of Sec.~\ref{sec:app_EIH}; the second the screening,
Jeans physics, self-energy, and dispersion of
Secs.~\ref{sec:benchmarks}, \ref{sec:app_DF}, and
\ref{sec:graviton_splitting}; the third the in-medium memory and graviton
splitting of Secs.~\ref{sec:app_memory} and
\ref{sec:graviton_splitting}. The rest of this subsection defines the
three replacements precisely; on a first reading, the takeaways are
Eqs.~\eqref{eq:T_gen}, \eqref{eq:Gfull},
and~\eqref{eq:GammaTTT_exchange}.

For a homogeneous background with four-velocity $U^\mu$ ($U^2=1$), we decompose
each momentum into parts along and transverse to the flow,
$k^\mu = (U\cdot k)\,U^\mu + \kperp^\mu$, using the transverse projector
$\Delta^{\mu\nu} \equiv \eta^{\mu\nu} - U^\mu U^\nu$ so that
$\kperp^\mu \equiv \Delta^\mu{}_{\nu} k^\nu$. We write the positive spatial
invariant as $K^2 \equiv -\kperp^2 \ge 0$, and for several momenta
$E_i \equiv U\cdot k_i$, $K_i^2 \equiv -k_{i\perp}^2$, and
$Q_{ij} \equiv k_{i\perp}\cdot k_{j\perp}$. In
the fluid rest frame $U^\mu=(1,\mathbf 0)$ these reduce to
$\kperp^\mu=(0,\mathbf k)$, $K^2=|\mathbf k|^2$, and
$Q_{ij}=-\,\mathbf k_i\cdot \mathbf k_j$.

\paragraph{1. Worldline vertices.}
The worldline action $\Spp$ generates all point-particle--graviton
couplings. We work in coordinate-time gauge, $v^\mu=(1,\mathbf v)$, and
decompose the metric perturbation into potential and radiation modes. We use $H_{\mu\nu}=2h^{\rm pot}_{\mu\nu}$, so $D_H=4D$.
Expanding $\Spp$ in powers of the potential graviton $H_{\mu\nu}$ and the
relative velocity gives
{\small
\begin{align}
    \mathcal{L}_{v^0}
    &= \frac{F}{2\Mpl} H_{00},
    \qquad
    \mathcal{L}_{v^1}
    = \frac{F}{\Mpl} H_{0i} v^i,
    \label{eq:Lv0}\\
    \mathcal{L}_{v^2}
    &= \frac{H_{00}}{4\Mpl}
       \Big[F v^2+F'\,(-u^2-v^2+2\mathbf u\cdot\mathbf v)\Big]
       +\frac{H_{ij}}{2\Mpl}
       \Big[F v^iv^j-F'\,(v^i-u^i)(v^j-u^j)\Big]
       -\frac{F}{8\Mpl^2}H_{00}^2,
    \label{eq:vertices_v2}
\end{align}
}
where $u^i$ is the spatial part of the background fluid velocity evaluated on
the worldline. At this order the entire medium dependence of the worldline
sector is carried by the matched coefficients $F$, $F'$, and by the local
background flow.

For radiation observables the linear coupling is packaged as
$S_{\rm pp}^{(1)}=-\tfrac{\kappa}{2}\int\dd{^4x}\,
h_{\mu\nu}\mathcal T^{\mu\nu}_{\rm pp}$, with the exact worldline stress
tensor
{\small
\begin{equation}
    \boxed{\;
    \mathcal{T}^{\mu\nu}_{\rm pp}(x)
    = -\int d\lambda\,
      \Biggl[
        (F-\gamma F')\,
        \frac{v^\mu v^\nu}{\sqrt{v^2}}
        +F'\left(
            \bigl(v^\mu U^\nu+U^\mu v^\nu\bigr)
            -\gamma\sqrt{v^2}\,U^\mu U^\nu
          \right)
      \Biggr]
      \delta^{(4)}(x-x(\lambda))\,,\;
    }
    \label{eq:T_gen}
\end{equation}
}
where $\gamma\equiv (v\cdot U)/\sqrt{v^2}$. Here $F=F(\gamma)$ and $F'=dF/d\gamma$. In the vacuum limit
$F\to -M$ and $F'\to0$, this reduces to the standard point-particle source.

\paragraph{2. Dressed propagator.}
In vacuum, the graviton propagator in de~Donder gauge is
\begin{equation}
    D_{\mu\nu;\alpha\beta}(k)
    =
    \frac{i\,P_{\mu\nu;\alpha\beta}}{k^2+i\epsilon},
    \qquad
    P_{\mu\nu;\alpha\beta}
    \equiv
    \frac{1}{2}
    \left(
        \eta_{\mu\alpha}\eta_{\nu\beta}
        + \eta_{\mu\beta}\eta_{\nu\alpha}
        - \eta_{\mu\nu}\eta_{\alpha\beta}
    \right).
    \label{eq:vacprop}
\end{equation}
For a relativistic perfect fluid it is convenient to separate the seagull and transverse
pieces of the polarization tensor from the universal sound-pole
contribution:
\begin{equation}
    \Pi^{\mu\nu\alpha\beta}(k)
    =
    \Pi^{\mu\nu\alpha\beta}_{\rm rest}(k)
    +
    \Pi^{\mu\nu\alpha\beta}_{s}(k),
    \qquad
    \Pi^{\mu\nu\alpha\beta}_{s}(k)
    \equiv
    \mathcal Z_s(k)\,
    t_s^{\mu\nu}(k)\,
    t_s^{\alpha\beta}(-k).
    \label{eq:Pi_fluid_full}
\end{equation}
Here $t_s^{\mu\nu}(k)$ is the linearized sound-mode stress-tensor form
factor, derived from the fluid EFT in App.~\ref{app:phonon_vertex}:
\begin{equation}
    t_s^{\mu\nu}(k)
    =
    (1+c_s^2)\,U^\mu U^\nu
    +\frac{U\cdot k}{K^2}\bigl(U^\mu k_\perp^\nu + U^\nu k_\perp^\mu\bigr)
    -c_s^2\,\eta^{\mu\nu},
    \label{eq:ts_explicit}
\end{equation}
normalized so that $t_s^{00}(k)=1$ in the fluid rest frame (where
$k_\perp^0=0$). The sound pole
is carried by
\begin{equation}
    \mathcal Z_s(k)
    =
    -\,\frac{(\epsilon+p)\,K^2}
         {(U\cdot k)^2-c_s^2 K^2+i\epsilon},
    \label{eq:Zs_def}
\end{equation}
whose overall minus sign follows from
$\Pi\equiv i\langle\delta T\,\delta T\rangle^{\rm conn}$ with a stable
longitudinal mode and fixes the sign of the dispersion shift in
Sec.~\ref{sec:graviton_splitting}. On the vacuum light cone $k^2=0$, where
$(U\cdot k)^2=K^2$, this reduces to
\begin{equation}
    \mathcal Z_s(k)\big|_{k^2=0}
    = -\,\frac{\epsilon+p}{1-c_s^2},
    \label{eq:Zs_onshell}
\end{equation}
negative for any stable fluid. The remainder is
\begin{equation*}
    \Pi^{\mu\nu\alpha\beta}_{\rm rest}
    =4\langle\mathcal S^{\mu\nu\alpha\beta}\rangle
    +w\bigl(U^\mu U^\alpha Q^{\nu\beta}+3\ \text{perms}\bigr),
    \qquad
    Q^{\mu\nu}=\Delta^{\mu\nu}+\frac{k_\perp^\mu k_\perp^\nu}{K^2}.
\end{equation*}
The second term is the transverse-displacement contribution; it has no sound
pole. Together with the seagull~\eqref{eq:seagull_fluid}, it completes the
off-shell kernel and its Ward identity~\eqref{eq:ward}.

The full propagator is obtained by Dyson resummation,
\begin{equation}
    \boxed{\;
    G_{\mu\nu;\alpha\beta}
    =
    \Bigl[
        D^{-1}
        - i\left(\frac{\kappa}{2}\right)^2 \Pi
    \Bigr]^{-1}_{\mu\nu;\alpha\beta}
    =
    D_{\mu\nu;\alpha\beta}
    + i D_{\mu\nu;\rho\sigma}
      \left(\frac{\kappa}{2}\right)^2
      \Pi^{\rho\sigma\gamma\delta}
      D_{\gamma\delta;\alpha\beta}
    + \cdots\,.\;
    }
    \label{eq:Gfull}
\end{equation}
This is the precise sense in which the medium leaves the diagram topology
unchanged: every vacuum internal graviton line is simply replaced by the
dressed propagator $G$.

For causal observables computed in the in-in formalism, the same replacement-rule
structure holds with $D$, $\Pi$, and $G$ replaced by the corresponding retarded
kernels. With the overall
Feynman $i$ stripped, $D_R=K_R^{-1}$ and
$G_R=[D_R^{-1}+(\kappa/2)^2\Pi_R]^{-1}$. The anti-Hermitian part of the retarded $\Pi$ is the stress-tensor
spectral density of the medium; that single object drives the dissipative
self-force and its renormalization-group flow in the companion
paper~\cite{Modrekiladze:2026twz}, while its Hermitian part controls the
conservative effects below.

\paragraph{3. Medium-induced cubic graviton vertex.}
The same logic produces a genuinely new interaction. Because the medium
responds nonlinearly to the metric, integrating it out endows the graviton with
a cubic self-coupling that has no vacuum analogue---the seed of the graviton
splitting $h\to hh$ of Sec.~\ref{sec:graviton_splitting}. The vertex is the
momentum-space transform of the connected three-point function of
Eq.~\eqref{eq:Seff_cubic},
$\Gamma^{\mu\nu,\alpha\beta,\rho\sigma}_{TTT}(k_1,k_2,k_3)
\equiv\langle\delta T^{\mu\nu}(k_1)\,\delta T^{\alpha\beta}(k_2)\,
\delta T^{\rho\sigma}(k_3)\rangle^{\rm conn}_{\varphi}$, with all momenta
incoming and $k_1+k_2+k_3=0$.

For a relativistic perfect fluid, the leading nonanalytic contribution is
carried by the longitudinal sound mode. Using the same linearized
sound-mode tensor $t_s^{\mu\nu}(k)$ of Eq.~\eqref{eq:Pi_fluid_full}, we
write
\begin{equation}
    \delta T^{\mu\nu}(k)= w\,t_s^{\mu\nu}(k)\,\psi(k),
    \qquad
    w\equiv \epsilon+p,
\end{equation}
with Feynman propagator
\begin{equation}
    G_F^{\psi}(k)
    =
    \frac{i\,K^2}
         {w\big[(U\cdot k)^2-c_s^2 K^2+i\epsilon\big]}.
    \label{eq:psi_prop}
\end{equation}
The sound-exchange contribution to $\Gamma_{TTT}$ then factorizes as
\begin{equation}
    \Gamma^{\mu\nu,\alpha\beta,\rho\sigma}_{TTT,\rm exch}(k_1,k_2,k_3)
    =
    w^3\,
    t_s^{\mu\nu}(k_1)\,
    t_s^{\alpha\beta}(k_2)\,
    t_s^{\rho\sigma}(k_3)\,
    G_F^\psi(k_1)\,G_F^\psi(k_2)\,G_F^\psi(k_3)\,
    i\,\mathcal V_\psi(k_1,k_2,k_3),
    \label{eq:GammaTTT_exchange}
\end{equation}
where $\mathcal V_\psi$ is the amputated longitudinal three-phonon vertex,
fixed by the cubic Lagrangian of the perfect-fluid
EFT~\cite{Endlich:2010hf} and derived in App.~\ref{app:phonon_vertex}. Here
we need only its structural form: it is polynomial in $(E_i,k_{i\perp})$,
manifestly $S_3$-symmetric, carries its universal nonanalytic content
entirely in the three external sound poles, and depends on the equation of
state only through $c_s^2$ and the cubic coefficient
$f_3\equiv w\,\partial c_s^2/\partial\epsilon+c_s^2(c_s^2-1)$. The full
kernel also receives contact contributions from the nonlinear part of the
fluid stress tensor,
$\Gamma_{TTT}=\Gamma_{TTT,\rm exch}+\Gamma_{TTT,\rm contact}$, defining the
exact in-medium cubic graviton vertex
$V^{(3)}_{\rm med}=V^{(3)}_{\rm EH}+V^{(3)}_\Pi$ with
$V^{(3)}_\Pi\equiv(\kappa/2)^3\,\Gamma_{TTT}$; Eq.~\eqref{eq:GammaTTT_exchange} gives the three-linear-stress exchange.
The $h\to hh$ amplitudes also require the nonlinear stress contacts.

This completes the in-medium Feynman rules. The remainder of this section
supplies context rather than machinery: Sec.~\ref{sec:media_catalogue}
catalogues how other media enter the same rules, Sec.~\ref{sec:benchmarks}
checks them against Jeans physics, and Sec.~\ref{sec:RG} discusses
matching and running. Readers interested in applications can proceed
directly to Sec.~\ref{sec:applications_nonzero}.

\subsection{A Catalogue of Media}
\label{sec:media_catalogue}

The construction of Sec.~\ref{sec:medium} never used any property of the
medium beyond its stress-energy correlators: the environment enters only
through $\Pi=i\langle\delta T\,\delta T\rangle^{\rm conn}$, the higher
vertices $\Gamma$, and the one-body data $F,F'$, so different media are
not different theories but different correlators inserted into the
\emph{same} diagrams. A convenient organizing quantity is the static
longitudinal response, which for any homogeneous medium defines a
screening (Jeans) wavenumber,
\begin{equation}
    \Pi^{0000}(k)\big|_{\rm static}
    \;\equiv\;
    \frac{k_J^2}{4\pi G_N},
    \label{eq:static_universal}
\end{equation}
below which the perturbative expansion in $\Pi$ insertions must be
resummed into the Poisson--Jeans system of Sec.~\ref{sec:benchmarks}. The
resulting Jeans scales, which bound the size of a stable homogeneous
background, were used as infrared cutoffs in the companion
paper~\cite{Modrekiladze:2026twz}; here we exhibit the response functions
behind them and the single replacement each medium makes.

\paragraph{Relativistic perfect fluid.}
This is the worked example of Sec.~\ref{sec:feynman}: a propagating sound
pole, $\mathcal Z_s$ of Eq.~\eqref{eq:Zs_def}, with
$\Pi^{0000}_{\rm static}=(\epsilon+p)/c_s^2$ and hence
$k_J^2=4\pi G_N(\epsilon+p)/c_s^2$ [Eq.~\eqref{eq:newton_pi_relative}];
pressure support ($c_s^2>0$) keeps the response finite.

\paragraph{Collisionless matter.}
A gas of collisionless particles with one-body phase-space distribution
$\bar f(\mathbf v)$, normalized to the mass density $\rho$, is governed not
by hydrodynamics but by the Vlasov--Poisson system. Linearizing it yields
the gravitational analogue of the Lindhard response~\cite{BinneyTremaine},
\begin{equation}
    \Pi^{0000}(\omega,\mathbf k)
    =
    \int d^3v\,
    \frac{\mathbf k\cdot\partial_{\mathbf v}\bar f(\mathbf v)}
         {\omega-\mathbf k\cdot\mathbf v+i\epsilon}\,,
    \label{eq:Pi_collisionless}
\end{equation}
which differs from the fluid in two universal ways. First, the sharp
sound pole is replaced by a branch cut over the support of $\bar f$: modes
whose phase velocity lies inside the velocity distribution are
Landau-damped, the microscopic origin of collisionless dynamical
friction~\cite{Chandrasekhar:1943ys}. Second, the static response is
regulated by the velocity dispersion rather than by pressure,
$\Pi^{0000}_{\rm static}=\rho/\sigma^2$ and $k_J^2=4\pi G_N\rho/\sigma^2$,
so $\sigma$ plays the role of $c_s$. The dust self-force of the companion
paper~\cite{Modrekiladze:2026twz} is the $\sigma\to0$ limit of this
kernel, in which the cut collapses to a pole; at finite $\sigma$, inserting
\eqref{eq:Pi_collisionless} into the dressed propagator \eqref{eq:Gfull}
yields the Landau-damped self-force, which has not been computed.

\paragraph{The dust point.}
Pressureless matter is the common singular limit: $c_s\to0$ (or
$\sigma\to0$) drives $k_J\to\infty$, so cold matter is Jeans-unstable on
all scales and admits no stable homogeneous background. A nonzero
$c_s^2$, a velocity dispersion $\sigma^2$, or, as we show next, a quantum
pressure must therefore regulate the deep infrared; the dust results of
the companion paper~\cite{Modrekiladze:2026twz} are obtained with
$L<\lambda_J$ held fixed while the regulator is sent to zero inside the
response only.

\paragraph{Coherent scalar field (wave dark matter).}
Consider a real scalar with
$\mathcal L=\tfrac12(\partial\phi)^2-\tfrac12 m^2\phi^2$ in a homogeneous
oscillating background $\phi_0(t)=\phi_a\cos(mt)$. The energy density
$\epsilon=\tfrac12 m^2\phi_a^2$ is constant while the pressure
$p(t)=-\epsilon\cos(2mt)$ oscillates with zero mean, so on time-scales long compared
with $m^{-1}$ the condensate is pressureless, like dust, and the infrared
is instead regulated by gradient (``quantum'') energy. In the regime
$\omega,|\mathbf k|\ll m$ relevant to gravitational-wave astronomy, the
dynamics reduces to the Schr\"odinger--Poisson (Madelung) system, and
density perturbations obey
$\omega^2=|\mathbf k|^4/4m^2-4\pi G_N\rho$~\cite{Hu:2000ke,Hui:2016ltb}%
---precisely the perfect-fluid sound pole carrying a
\emph{momentum-dependent} sound speed,
\begin{equation}
    \boxed{\;c_s^2(\mathbf k)=\frac{|\mathbf k|^2}{4m^2}\,,\qquad |\mathbf k|\ll m\,.\;}
    \label{eq:cs_scalar}
\end{equation}
The medium-induced polarization is then the fluid form
\eqref{eq:Zs_def} under this replacement,
\begin{equation}
    \mathcal Z_s^{(\phi)}(k)
    =
    -\,\frac{\epsilon\,K^2}
         {(U\cdot k)^2-\dfrac{K^4}{4m^2}+i\epsilon}\,,
    \label{eq:Zs_scalar}
\end{equation}
whose rest-frame pole is the de~Broglie dispersion $\omega=\mathbf k^2/2m$.
The static response is itself momentum dependent,
$\Pi^{0000}_{\rm static}\propto m^2/\mathbf k^2$, so the screening scale is
fixed self-consistently by the marginal mode of the quartic dispersion,
giving the quantum-Jeans wavenumber $k_J=(16\pi G_N\rho\,m^2)^{1/4}$. Two
consequences follow from the replacement rule: the black-hole Love-number
resonance $\omega=c_s K$ of Sec.~\ref{sec:app_love} migrates to
$\omega=\mathbf k^2/2m$, the bound-state spectrum of a ``gravitational
atom''~\cite{Brito:2015oca,Baumann:2018vus}; and because
$c_s^2(\mathbf k)\ll1$, 
the two-quantum continuum that dresses the transverse-traceless graviton
extends across the light cone. A feature with no fluid analogue is
the oscillating background itself, whose $2m$ harmonic we exploit in
App.~\ref{app:benchmarks}.

\begin{table}[!htb]
\centering
\small
\begin{tabular}{@{}l l l l p{3.5cm}@{}}
\toprule
Medium & IR regulator & Response pole & $k_J^2$ & Distinctive phenomenon \\
\midrule
Perfect fluid & $c_s^2$ & sound, $\omega=c_s|\mathbf k|$
  & $\dfrac{4\pi G_N(\epsilon+p)}{c_s^2}$
  &  Cherenkov friction; opacity from the sound continuum \\[0.4ex]
Collisionless gas & $\sigma^2$ & Landau cut
  & $\dfrac{4\pi G_N\rho}{\sigma^2}$
  & dynamical friction \\[0.4ex]
Coherent scalar & $\dfrac{|\mathbf k|^2}{4m^2}$ & de~Broglie, $\omega=\dfrac{\mathbf k^2}{2m}$
  & $4m\sqrt{\pi G_N\rho}$
  & gravitational-atom Love resonance; $2m$ line \\
\bottomrule
\end{tabular}
\caption{Three media as three choices of stress-energy correlator in the
same Feynman rules: the infrared regulator of the static response
\eqref{eq:static_universal}, the analytic structure of the longitudinal
response, and the screening (Jeans) wavenumber. Dust is the common singular
limit $k_J\to\infty$.}
\label{tab:media}
\end{table}

\subsection{Benchmarks: Jeans Physics}
\label{sec:benchmarks}

A framework that reorganizes known physics must return that physics on
demand. Resumming $\Pi$ insertions on the Newtonian line, each
contributing the relative factor $k_J^2/|\mathbf k|^2$ of
Eq.~\eqref{eq:newton_pi_relative}, turns the static potential into the
modified Poisson equation $(\nabla^2+k_J^2)\Phi=4\pi G_N\rho_{\rm ext}$,
and restoring the frequency dependence of the sound pole gives the dressed
longitudinal graviton the relativistic Jeans--sound dispersion
\begin{equation}
    \boxed{\;
    \omega^2=c_s^2|\mathbf k|^2-4\pi G_N(\epsilon+p)\,,
    \;}
    \label{eq:jeans_dispersion}
\end{equation}
the textbook Jeans relation for non-relativistic
matter~\cite{BinneyTremaine}, with the acoustic branch and the
gravitational instability emerging with their relative coefficient fixed,
not fitted. Specializing $c_s^2$ reproduces the classical,
quantum~\cite{Hu:2000ke,Hui:2016ltb}, and kinetic~\cite{BinneyTremaine}
Jeans criteria of Table~\ref{tab:media} from the same dressed propagator,
each by a single replacement of the correlator---the central consistency
check of the construction. The derivations, together with three
closed-form wave-dark-matter observables that follow from the same
propagator (the oscillatory two-body potential beyond the de~Broglie
scale, the Khmelnitsky--Rubakov pulsar-timing
potential~\cite{Khmelnitsky:2013lxt}, and gravitational-wave dispersion
through a condensate~\cite{Flauger:2017gtb}), are collected in
App.~\ref{app:benchmarks}.

\subsection{Matching, Renormalization, and the Renormalization Group}
\label{sec:RG}

The Wilson coefficients $F(1)$ and $F'(1)$ are not determined by the
long-distance EFT itself: they are matched once in the single-body problem
and then inserted into many-body calculations, with higher-derivative
operators, tidal couplings, and higher correlators entering at subleading
order.

\paragraph{One-body coefficients across media.}
The matched values are themselves medium-dependent, and comparing them
across the catalogue of Sec.~\ref{sec:media_catalogue} reveals a sharp
criterion. For the relativistic fluid, $F(1)=-M+\rho V_p$ and
$F'(1)=-\tfrac32\rho V_p$ [Eq.~\eqref{eq:F1}] are
\emph{buoyant}: the $\rho V_p$ pieces are the Archimedean response of a body
displacing a volume $V_p$ of fluid, and exist only because the fluid exerts
a pressure-gradient contact force, $\mathbf F_{\rm buoy}=-V_p\,\nabla p$, on
that volume. Media without such a force lack them: pressureless dust and a
collisionless gas exert no Archimedean push---collisionless matter
free-streams rather than being displaced---so
\begin{equation}
    F(1)=-M,\qquad F'(1)=0
    \qquad\text{(dust, collisionless matter).}
    \label{eq:FFp_dust}
\end{equation}
A coherent scalar field coupled to the compact object only through gravity
is the same: gravity is universal rather than a contact interaction, so
nothing is displaced and again $F(1)=-M$, $F'(1)=0$; the exception is a
body carrying a direct scalar charge, a boson star built from the same
field. For these media the entire environmental effect resides in the
bulk correlators $\Pi,\Gamma$; the nonzero $F'$ of the fluid, and the
conservative drag and Stokes law it controls
(Secs.~\ref{sec:app_EIH} and \ref{sec:app_DF}), are specific to media with
pressure support.

As in vacuum worldline EFT, UV sensitivity arises because compact objects
are treated as pointlike. The short-distance part of any in-medium diagram
is therefore local and is absorbed into operators already allowed by
symmetry on the worldline or in the bulk effective action, while
nonanalytic dependence on momenta or frequencies is long-distance and
encodes genuine propagation through the medium. This distinction recurs
throughout the applications below---most cleanly in the self-energy of
Sec.~\ref{sec:app_DF}: analytic terms renormalize Wilson coefficients,
while the poles and cuts of the dressed propagator and bulk vertices carry
the physical environmental response.

\paragraph{Classical running in the worldline EFT.}
This matching data is not static: the worldline coefficients run. In vacuum
this is the well-understood \emph{classical} renormalization-group flow of
the worldline EFT~\cite{Goldberger:2004jt,Kol:2007bc,Porto:2016pyg}, whose
$\hbar$-independent beta functions reproduce the Blanchet--Damour hereditary
(tail) logarithms; the canonical example is the running of the radiating
mass quadrupole,
\begin{equation}
    \mu\frac{d}{d\mu}I_{ij}(\omega,\mu)
    = -\frac{214}{105}\,(G_N M\omega)^2\,I_{ij}(\omega,\mu),
    \label{eq:tail_RG}
\end{equation}
the tail anomalous dimension of Goldberger and
Ross~\cite{Goldberger:2009qd,Goldberger:2012kf,Galley:2015kus}. Because the
medium leaves the diagram topologies untouched, this structure persists in
the in-medium theory.

\paragraph{From protected to running: in-medium Love numbers.}
The medium also changes the renormalization of the \emph{tidal} sector in a
sharp, symmetry-based way. In vacuum the static tidal response of a black
hole vanishes identically and does not run: a hidden
$\mathrm{SL}(2,\mathbb R)$ ``Love symmetry'' protects the
coefficient~\cite{Hui:2021vcv,Hui:2022vbh,Charalambous:2021mea,Charalambous:2021kcz}.
The medium-induced tidal coupling of Sec.~\ref{sec:app_love} is frequency
dependent by construction---the sound pole introduces the explicit scale
$\omega^2-c_s^2K^2$ that breaks the ladder structure---and therefore belongs
to the \emph{dynamical}, post-adiabatic class, which is unprotected and runs

logarithmically. The $-214/105$ of Eq.~\eqref{eq:tail_RG} is the anomalous
dimension of the radiating quadrupole, a different operator, and the paper does
not derive a tidal one; the coefficient is matched in the companion
analysis~\cite{Modrekiladze:2025love}. The environment thus converts a
symmetry-protected, non-running vacuum coefficient into a genuine
renormalization-group observable: the in-medium Love number is generated,
is frequency dependent, and flows. The spectrum of induced couplings is
also richer than in vacuum. A moving medium supplies a preferred direction
$U^\mu$, so the induced response is not a single isotropic scalar
$\alpha_E$: the flow splits it into components along and across $U^\mu$,
mixes the electric and magnetic sectors, and adds a dissipative imaginary
part from the sound pole; a medium with vorticity or chirality further
induces parity-violating $E\!\cdot\!B$ couplings of the type analyzed
in~\cite{Modrekiladze:2022ioh}. Each of these vanishes for an isolated
black hole, so any one of them is a smoking gun for an environment.
\section{Applications I: Topologies with Nonzero Vacuum Limits}
\label{sec:applications_nonzero}

We now promote standard vacuum calculations to the in-medium case,
beginning with topologies that are nonzero already in vacuum GR.


\subsection{Einstein--Infeld--Hoffmann Potential}
\label{sec:app_EIH}

The 1PN conservative dynamics of a binary in vacuum, first obtained by
Einstein, Infeld, and Hoffmann~\cite{Einstein:1938yz}, was recast in the
worldline EFT of~\cite{Goldberger:2004jt}. We now compute its
generalization to a relativistic fluid background.

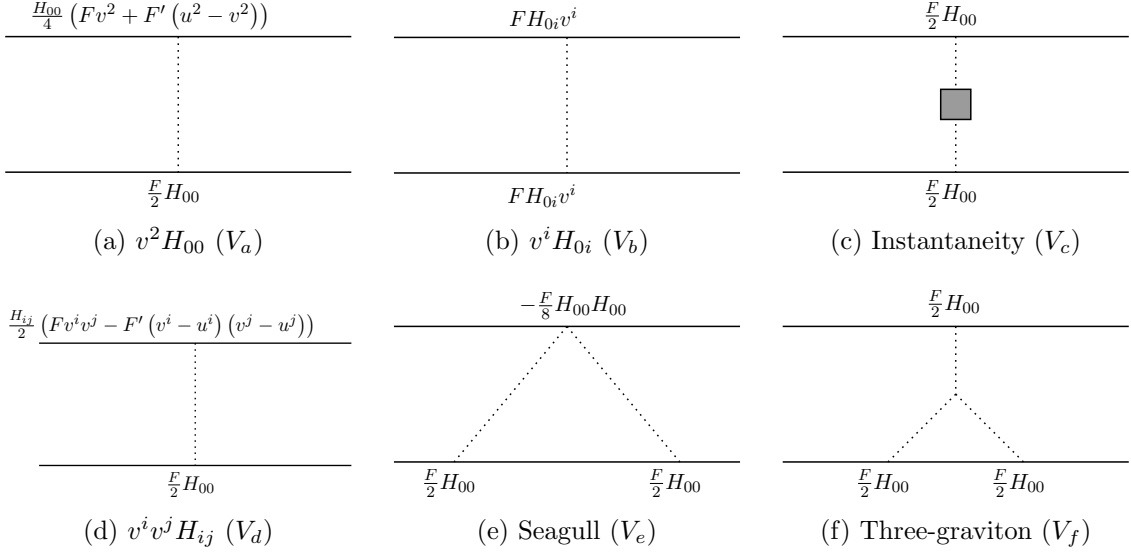
\begin{figure}[htbp]
    \centering
\tikzset{every picture/.style={line width=0.75pt}}
    \begin{subfigure}[b]{0.32\textwidth}
        \centering
\resizebox{0.95\linewidth}{!}{%
\begin{tikzpicture}[x=0.75pt,y=0.75pt,yscale=-1,xscale=1]
\draw    (235,110) -- (465,110) ;
\draw    (235,200) -- (465,200) ;
\draw  [dash pattern={on 0.84pt off 2.51pt}]  (350,110) -- (350,200) ;
\draw (327,203.4) node [anchor=north west][inner sep=0.75pt]    {$\frac{F}{2} H_{00}$};
\draw (250,85.4) node [anchor=north west][inner sep=0.75pt]    {$\frac{H_{00}}{4}\left( Fv^{2} +F'\left( u^{2} -v^{2}\right)\right)$};
\end{tikzpicture}}
        \caption{$v^2 H_{00}$ ($V_a$)}
    \end{subfigure}\hfill
    \begin{subfigure}[b]{0.32\textwidth}
        \centering
\resizebox{0.95\linewidth}{!}{%
\begin{tikzpicture}[x=0.75pt,y=0.75pt,yscale=-1,xscale=1]
\draw    (235,91) -- (465,91) ;
\draw    (235,181) -- (465,181) ;
\draw  [dash pattern={on 0.84pt off 2.51pt}]  (350,91) -- (350,181) ;
\draw (310,69.4) node [anchor=north west][inner sep=0.75pt]    {$F H_{0i} v^{i}$};
\draw (309,187.4) node [anchor=north west][inner sep=0.75pt]    {$F H_{0i} v^{i}$};
\end{tikzpicture}}
        \caption{$v^i H_{0i}$ ($V_b$)}
    \end{subfigure}\hfill
    \begin{subfigure}[b]{0.32\textwidth}
        \centering
\resizebox{0.95\linewidth}{!}{%
\begin{tikzpicture}[x=0.75pt,y=0.75pt,yscale=-1,xscale=1]
\draw    (235,110) -- (465,110) ;
\draw    (235,200) -- (465,200) ;
\draw  [dash pattern={on 0.84pt off 2.51pt}]  (350,110) -- (350,200) ;
\draw  [fill={rgb, 255:red, 155; green, 155; blue, 155 }  ,fill opacity=1 ] (340,145) -- (360,145) -- (360,165) -- (340,165) -- cycle ;
\draw (327,203.4) node [anchor=north west][inner sep=0.75pt]    {$\frac{F}{2} H_{00}$};
\draw (327,85.4) node [anchor=north west][inner sep=0.75pt]    {$\frac{F}{2} H_{00}$};
\end{tikzpicture}}
        \caption{Instantaneity ($V_c$)}
    \end{subfigure}

    \vspace{2ex}
    \begin{subfigure}[b]{0.32\textwidth}
        \centering
\resizebox{0.95\linewidth}{!}{%
\begin{tikzpicture}[x=0.75pt,y=0.75pt,yscale=-1,xscale=1]
\draw    (235,110) -- (465,110) ;
\draw    (235,200) -- (465,200) ;
\draw  [dash pattern={on 0.84pt off 2.51pt}]  (350,110) -- (350,200) ;
\draw (327,203.4) node [anchor=north west][inner sep=0.75pt]    {$\frac{F}{2} H_{00}$};
\draw (210,84.4) node [anchor=north west][inner sep=0.75pt]    {$\frac{H_{ij}}{2}\left( Fv^{i} v^{j} -F'\left( v^{i} -u^{i}\right)\left( v^{j} -u^{j}\right)\right)$};
\end{tikzpicture}}
        \caption{$v^iv^j H_{ij}$ ($V_d$)}
    \end{subfigure}\hfill
    \begin{subfigure}[b]{0.32\textwidth}
        \centering
\resizebox{0.95\linewidth}{!}{%
\begin{tikzpicture}[x=0.75pt,y=0.75pt,yscale=-1,xscale=1]
\draw    (235,110) -- (465,110) ;
\draw    (235,200) -- (465,200) ;
\draw  [dash pattern={on 0.84pt off 2.51pt}]  (350,110) -- (275,200) ;
\draw  [dash pattern={on 0.84pt off 2.51pt}]  (350,110) -- (425,200) ;
\draw (252,203.4) node [anchor=north west][inner sep=0.75pt]    {$\frac{F}{2} H_{00}$};
\draw (317,85.4) node [anchor=north west][inner sep=0.75pt]    {$-\frac{F}{8} H_{00} H_{00}$};
\draw (402,203.4) node [anchor=north west][inner sep=0.75pt]    {$\frac{F}{2} H_{00}$};
\end{tikzpicture}}
        \caption{Seagull ($V_e$)}
    \end{subfigure}\hfill
    \begin{subfigure}[b]{0.32\textwidth}
        \centering
\resizebox{0.95\linewidth}{!}{%
\begin{tikzpicture}[x=0.75pt,y=0.75pt,yscale=-1,xscale=1]
\draw    (235,110) -- (465,110) ;
\draw    (235,200) -- (465,200) ;
\draw  [dash pattern={on 0.84pt off 2.51pt}]  (350,155) -- (305,200) ;
\draw  [dash pattern={on 0.84pt off 2.51pt}]  (350,155) -- (395,200) ;
\draw  [dash pattern={on 0.84pt off 2.51pt}]  (350,110) -- (350,155) ;
\draw (282,203.4) node [anchor=north west][inner sep=0.75pt]    {$\frac{F}{2} H_{00}$};
\draw (329,84.4) node [anchor=north west][inner sep=0.75pt]    {$\frac{F}{2} H_{00}$};
\draw (372,203.4) node [anchor=north west][inner sep=0.75pt]    {$\frac{F}{2} H_{00}$};
\end{tikzpicture}}
        \caption{Three-graviton ($V_f$)}
    \end{subfigure}
    \caption{Feynman diagrams contributing to the 1PN conservative
    potential in a fluid background. Topologies are identical to the
    vacuum EIH calculation; the medium enters only through the matched
    coefficients $F_i,F_i'$.}
    \label{fig:EIH_diagrams}
\end{figure}

The 1PN conservative potential arises from the same six potential-mode
topologies as in the vacuum EIH calculation; see
Fig.~\ref{fig:EIH_diagrams}. Because potential modes carry $k_0=0$, the
leading conservative medium dependence at this order is encoded in the
matched worldline coefficients $F_i,F_i'$. We therefore evaluate the
exchange graphs with the instantaneous vacuum potential propagator and use
the generalized vertices~\eqref{eq:Lv0}--\eqref{eq:vertices_v2}. Explicit
potential-region insertions of the bulk polarization tensor can be
incorporated within the same replacement-rule framework, but are not needed
for the result displayed here.

The evaluation of the six graphs parallels the vacuum calculation
step by step, with $V_0\equiv-G_NF_1F_2/r$ and
$\mathbf r\equiv\mathbf x_1-\mathbf x_2$: the gravito-magnetic ($V_b$),
instantaneity ($V_c$), and nonlinear ($V_e,V_f$) topologies depend only on
$F_i$, while the drag coefficients $F_i'$ enter only through the $H_{00}$
and $H_{ij}$ vertices of $V_a$ and $V_d$. Summing all contributions yields
the compact 1PN conservative potential
\begin{align}
V_{\rm EIH} =\;&
\frac{G_N^2}{2r^2}\,F_1F_2(F_1+F_2)
\notag\\
&+\frac{V_0}{2}\left[
-3(v_1^2+v_2^2)
+7(\mathbf v_1\cdot \mathbf v_2)
+\frac{(\mathbf v_1\cdot \mathbf r)(\mathbf v_2\cdot \mathbf r)}{r^2}
\right]
\notag\\
&+\frac{V_0}{2}
\left[
\frac{F_1'}{F_1}\bigl(3v_1^2-6\mathbf v_1\cdot \mathbf u_1+3u_1^2\bigr)
+\frac{F_2'}{F_2}\bigl(3v_2^2-6\mathbf v_2\cdot \mathbf u_2+3u_2^2\bigr)
\right].
\label{eq:VEIH_compact}
\end{align}
Here and below, $V_{\rm EIH}$ is quoted with the sign convention in which
it enters the two-body Lagrangian, $L\supset V_{\rm EIH}$---the object the
diagrams compute directly; the Hamiltonian potential follows by the
standard Legendre transform. This is the vacuum EIH result with the mass
insertions replaced by the matched coefficients $F_i$, supplemented by the
rigidly patterned medium correction proportional to $F_i'$.

Writing $\delta_i\equiv\rho_i V_{p,i}$, so that $F_i=-M_i+\delta_i$ and
$F_i'=+\tfrac{3}{2}\delta_i$, the linear-density expansion becomes
\begin{align}
V_{\rm EIH}
=\;&V_{\rm EIH}^{\rm vac}(M_1,M_2)
\notag\\
&+\frac{G_N^2}{2r^2}
\Big[
\delta_1 M_2(2M_1+M_2)
+\delta_2 M_1(M_1+2M_2)
\Big]
\notag\\
&-\frac{G_N}{2r}
\Big(M_2\delta_1+M_1\delta_2\Big)
\left[
3(v_1^2+v_2^2)
-7(\mathbf v_1\cdot \mathbf v_2)
-\frac{(\mathbf v_1\cdot \mathbf r)(\mathbf v_2\cdot \mathbf r)}{r^2}
\right]
\notag\\
&+\frac{3G_N}{4r}
\Big[
M_2\delta_1\bigl(3v_1^2-6\mathbf v_1\cdot \mathbf u_1+3u_1^2\bigr)
+M_1\delta_2\bigl(3v_2^2-6\mathbf v_2\cdot \mathbf u_2+3u_2^2\bigr)
\Big],
\label{eq:VEIH_full}
\end{align}
where
\begin{equation}
    V_{\rm EIH}^{\rm vac}
    =
    -\frac{G_N^2}{2r^2}M_1M_2(M_1+M_2)
    +\frac{G_N M_1M_2}{2r}
    \left[
        3(v_1^2+v_2^2)
        -7(\mathbf v_1\cdot \mathbf v_2)
        -\frac{(\mathbf v_1\cdot \mathbf r)(\mathbf v_2\cdot \mathbf r)}{r^2}
    \right].
\end{equation}
Here $\delta_i$ and $\mathbf u_i$ are evaluated locally on worldline $i$.

\subsubsection{Three-Body Potentials from Fluid Mediation}

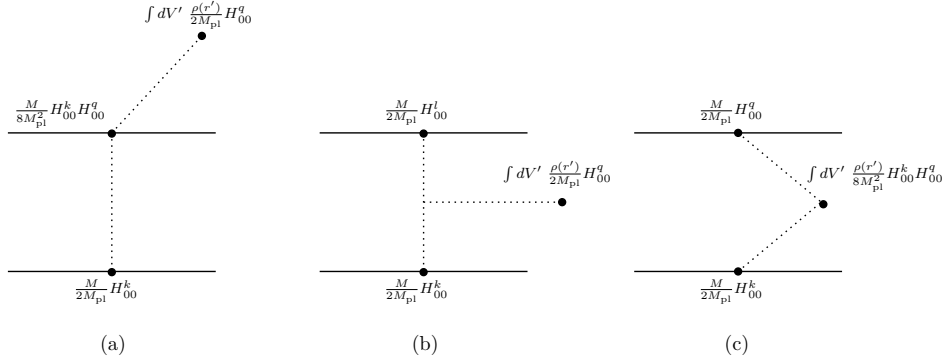
\begin{figure}[htbp]
\centering
\tikzset{every picture/.style={line width=0.75pt}}
\resizebox{0.82\textwidth}{!}{%
\begin{tikzpicture}[x=0.75pt,y=0.75pt,yscale=-1,xscale=1]
\draw    (5,129) -- (155,129) ;
\draw    (5,229) -- (155,229) ;
\draw  [dash pattern={on 0.84pt off 2.51pt}]  (80,129) -- (80,229) ;
\draw  [dash pattern={on 0.84pt off 2.51pt}]  (80,129) -- (145,59) ;
\draw    (230,129) -- (380,129) ;
\draw    (380,229) -- (230,229) ;
\draw  [dash pattern={on 0.84pt off 2.51pt}]  (305,129) -- (305,160.44) -- (305,229) ;
\draw  [dash pattern={on 0.84pt off 2.51pt}]  (305,179) -- (405,179) ;
\draw    (457,129) -- (607,129) ;
\draw    (607,229) -- (457,229) ;
\draw  [dash pattern={on 0.84pt off 2.51pt}]  (532,129) -- (592,179) ;
\draw  [dash pattern={on 0.84pt off 2.51pt}]  (532,229) -- (592,179) ;
\draw  [fill={rgb, 255:red, 0; green, 0; blue, 0 }  ,fill opacity=1 ] (77.5,229.5) .. controls (77.5,228.12) and (78.62,227) .. (80,227) .. controls (81.38,227) and (82.5,228.12) .. (82.5,229.5) .. controls (82.5,230.88) and (81.38,232) .. (80,232) .. controls (78.62,232) and (77.5,230.88) .. (77.5,229.5) -- cycle ;
\draw  [fill={rgb, 255:red, 0; green, 0; blue, 0 }  ,fill opacity=1 ] (77.5,129.5) .. controls (77.5,128.12) and (78.62,127) .. (80,127) .. controls (81.38,127) and (82.5,128.12) .. (82.5,129.5) .. controls (82.5,130.88) and (81.38,132) .. (80,132) .. controls (78.62,132) and (77.5,130.88) .. (77.5,129.5) -- cycle ;
\draw  [fill={rgb, 255:red, 0; green, 0; blue, 0 }  ,fill opacity=1 ] (302.5,129.5) .. controls (302.5,128.12) and (303.62,127) .. (305,127) .. controls (306.38,127) and (307.5,128.12) .. (307.5,129.5) .. controls (307.5,130.88) and (306.38,132) .. (305,132) .. controls (303.62,132) and (302.5,130.88) .. (302.5,129.5) -- cycle ;
\draw  [fill={rgb, 255:red, 0; green, 0; blue, 0 }  ,fill opacity=1 ] (302.5,229.5) .. controls (302.5,228.12) and (303.62,227) .. (305,227) .. controls (306.38,227) and (307.5,228.12) .. (307.5,229.5) .. controls (307.5,230.88) and (306.38,232) .. (305,232) .. controls (303.62,232) and (302.5,230.88) .. (302.5,229.5) -- cycle ;
\draw  [fill={rgb, 255:red, 0; green, 0; blue, 0 }  ,fill opacity=1 ] (402.5,179) .. controls (402.5,177.62) and (403.62,176.5) .. (405,176.5) .. controls (406.38,176.5) and (407.5,177.62) .. (407.5,179) .. controls (407.5,180.38) and (406.38,181.5) .. (405,181.5) .. controls (403.62,181.5) and (402.5,180.38) .. (402.5,179) -- cycle ;
\draw  [fill={rgb, 255:red, 0; green, 0; blue, 0 }  ,fill opacity=1 ] (591,180.5) .. controls (591,179.12) and (592.12,178) .. (593.5,178) .. controls (594.88,178) and (596,179.12) .. (596,180.5) .. controls (596,181.88) and (594.88,183) .. (593.5,183) .. controls (592.12,183) and (591,181.88) .. (591,180.5) -- cycle ;
\draw  [fill={rgb, 255:red, 0; green, 0; blue, 0 }  ,fill opacity=1 ] (529.5,129) .. controls (529.5,127.62) and (530.62,126.5) .. (532,126.5) .. controls (533.38,126.5) and (534.5,127.62) .. (534.5,129) .. controls (534.5,130.38) and (533.38,131.5) .. (532,131.5) .. controls (530.62,131.5) and (529.5,130.38) .. (529.5,129) -- cycle ;
\draw  [fill={rgb, 255:red, 0; green, 0; blue, 0 }  ,fill opacity=1 ] (529.5,229) .. controls (529.5,227.62) and (530.62,226.5) .. (532,226.5) .. controls (533.38,226.5) and (534.5,227.62) .. (534.5,229) .. controls (534.5,230.38) and (533.38,231.5) .. (532,231.5) .. controls (530.62,231.5) and (529.5,230.38) .. (529.5,229) -- cycle ;
\draw  [fill={rgb, 255:red, 0; green, 0; blue, 0 }  ,fill opacity=1 ] (142.5,59) .. controls (142.5,57.62) and (143.62,56.5) .. (145,56.5) .. controls (146.38,56.5) and (147.5,57.62) .. (147.5,59) .. controls (147.5,60.38) and (146.38,61.5) .. (145,61.5) .. controls (143.62,61.5) and (142.5,60.38) .. (142.5,59) -- cycle ;
\draw (51, 233.4) node [anchor=north west][inner sep=0.75pt]  [font=\scriptsize]  {$\frac{M}{2M_{\rm pl}} H{_{00}^{k}}$};
\draw (361,147.4) node [anchor=north west][inner sep=0.75pt]  [font=\scriptsize]  {$\int dV'\ \frac{\rho ( r')}{2M_{\rm pl}} H{_{00}^{q}}$};
\draw (8,104.4) node [anchor=north west][inner sep=0.75pt]  [font=\scriptsize]  {$\frac{M}{8M_{\rm pl}^{2}} H_{00}^{k} H{_{00}^{q}}$};
\draw (275,233.4) node [anchor=north west][inner sep=0.75pt]  [font=\scriptsize]  {$\frac{M}{2M_{\rm pl}} H{_{00}^{k}}$};
\draw (275,104.4) node [anchor=north west][inner sep=0.75pt]  [font=\scriptsize]  {$\frac{M}{2M_{\rm pl}} H{_{00}^{l}}$};
\draw (102,32.4) node [anchor=north west][inner sep=0.75pt]  [font=\scriptsize]  {$\int dV'\ \frac{\rho ( r')}{2M_{\rm pl}} H{_{00}^{q}}$};
\draw (580,147.4) node [anchor=north west][inner sep=0.75pt]  [font=\scriptsize]  {$\int dV'\ \frac{\rho ( r')}{8M_{\rm pl}^{2}} H{_{00}^{k}} H{_{00}^{q}}$};
\draw (502,233.4) node [anchor=north west][inner sep=0.75pt]  [font=\scriptsize]  {$\frac{M}{2M_{\rm pl}} H{_{00}^{k}}$};
\draw (502,104.4) node [anchor=north west][inner sep=0.75pt]  [font=\scriptsize]  {$\frac{M}{2M_{\rm pl}} H{_{00}^{q}}$};
\draw (70,273) node [anchor=north west][inner sep=0.75pt]   [align=left] {(a)};
\draw (520.67,273) node [anchor=north west][inner sep=0.75pt]   [align=left] {(c)};
\draw (295,273) node [anchor=north west][inner sep=0.75pt]   [align=left] {(b)};
\end{tikzpicture}%
}
\caption{Three-body potentials from fluid-mediated graviton exchange at 1PN.}
\label{fig:3body}
\end{figure}

Complete waveform templates also require the gravitational interaction
between the binary and the surrounding fluid. At 1PN order the fluid
mediates three-body interactions between the compact objects and a fluid
element at $\mathbf x_3$, with three distinct topologies
(Fig.~\ref{fig:3body}); these vanish identically in vacuum and have no
analog in modified gravity theories. Because these topologies are already
first order in the ambient density, the compact-body legs may be
evaluated with their vacuum masses $M_i$.

The three topologies of Fig.~\ref{fig:3body} evaluate to
\begin{align}
    V_{2a} &= -\frac{M_1M_2}{64\Mpl^4}\frac{1}{4\pi r}\int dV'\frac{\rho(r')}{4\pi r'}
    = -\tfrac12 V_{2b}, \\
    V_{2c} &= -\frac{M_1M_2}{64\Mpl^4}\int dV'\,\rho(r')\frac{1}{4\pi|\mathbf{r}'-\mathbf{r}_1|}\frac{1}{4\pi|\mathbf{r}'-\mathbf{r}_2|},
\end{align}
where $\mathbf{r}=\mathbf{x}_1-\mathbf{x}_2$ and
$\mathbf{r}'=\mathbf{x}_1-\mathbf{x}_3$; topology (b) equals $-2$ times
topology (a).

\paragraph{1PN correction to the Stokes drag.}
The classical Stokes drag also receives a 1PN correction, derived from
the worldline action of~\cite{Modrekiladze:2024htc} in the Keldysh
formalism; to our knowledge this is the first systematic EFT derivation
of the relativistic Stokes corrections. To $\mathcal O(v^2/c^2)$:
\begin{equation}
    \mathbf{F}_{\rm drag}
    \approx 6\pi\rho\nu R\!\left[(\mathbf{u}-\mathbf{v})
    \!\left(1+\frac{u^2-v^2}{2}\right)
    +\mathbf{v}\bigl(\mathbf{v}\cdot(\mathbf{u}-\mathbf{v})\bigr)\right]
    +\tfrac{1}{2}K'(1)|\mathbf{u}-\mathbf{v}|^2(\mathbf{u}-\mathbf{v}),
    \label{eq:Stokes1PN}
\end{equation}
where $\nu$ is the kinematic viscosity and $K'(1)$ is the derivative of
the next dissipative Wilson coefficient $K(\hat\gamma)$ appearing in the
worldline EFT at order $(\hat\gamma-1)^2$ (not to be confused with the
momentum invariant $K$ of Sec.~\ref{sec:feynman}).

Together with the conservative potentials above, these supply the dynamical
input for waveform templates in relativistic environments; finite-size spin
corrections can be incorporated following Porto and
Rothstein~\cite{Porto:2005ac,Porto:2006xv}.

\subsubsection{Bulk Propagator and Cubic Vertex on EIH Topologies}
\label{sec:Newton_Pi}

The 1PN result \eqref{eq:VEIH_full} uses the vacuum potential propagator
and cubic vertex on every internal line. We now check the contributions of
the remaining in-medium building blocks---the dressed propagator
\eqref{eq:Gfull} and the medium cubic vertex of
Eq.~\eqref{eq:GammaTTT_exchange}---on the leading Newton exchange (the
$\mathcal O(v^0)$ limit of Fig.~\ref{fig:EIH_diagrams}(a), with both
vertices $\tfrac{F_i}{2}H_{00}$) and on the three-graviton topology of
Fig.~\ref{fig:EIH_diagrams}(f).

\paragraph{Static $\Pi^{0000}$ and the role of $\epsilon_{\rm env}$.}
A bulk-$\Pi$ insertion on the internal line of this Newton exchange
brings a factor
$(\kappa/2)^2\Pi\sim G_N(\epsilon+p)$, superficially a $G_N^2$ (1PN-sized)
contribution. But the extra $G_N$ does not stand alone: it combines with
$(\epsilon+p)L^2$, where $L$ is the IR scale of the diagram, into the
dimensionless Jeans ratio $\epsilon_{\rm env}/c_s^2$; this ratio, not
loose $G_N$ counting, controls bulk-propagator dressing. To verify this, note
that both vertices of the Newton exchange are $\frac{F_i}{2}H_{00}$
[Eq.~\eqref{eq:Lv0}], so what enters is $\Pi^{0000}(k)$ in the static
limit $k^0\to 0$. Since Eqs.~\eqref{eq:Pi_fluid_full}, \eqref{eq:Zs_def},
and \eqref{eq:ts_explicit} give $t_s^{00}(k)=1$ exactly,
\begin{equation}
    \boxed{\;
    \Pi^{0000}(k)\big|_{\rm static}
    \;=\;\frac{\epsilon+p}{c_s^2}\,{}+\epsilon\,.\;
    }
    \label{eq:Pi00_static}
\end{equation}

This is independent of $\mathbf k$: the static density--density response
is local, characterized by the Jeans combination $(\epsilon+p)/c_s^2$. The
relative correction to the momentum-space exchange amplitude from one
$\Pi$ insertion is
\begin{equation}
    \frac{\delta\mathcal A^{(\Pi)}}{\mathcal A_{\rm Newton}}(\mathbf k)
    \;=\;\frac{(\kappa/2)^2 P_{0000}\,\Pi^{0000}}{|\mathbf k|^2}
    \;=\;\frac{k_J^2}{|\mathbf k|^2},
    \qquad
    k_J^2\equiv\frac{4\pi G_N(\epsilon+p)}{c_s^2},
    \label{eq:newton_pi_relative}
\end{equation}
where $P_{0000}=\tfrac12$ is the de~Donder weight carried by the $00$--$00$
graviton propagator attaching to the insertion (exactly
$\tfrac12(1+3c_s^2)^2$ per insertion, reducing to $\tfrac12$
non-relativistically).
Resumming these insertions places the pole of the static potential at
$|\mathbf k|^2=k_J^2$---the linearized Poisson--Jeans result of
Sec.~\ref{sec:benchmarks}; the IR sensitivity of the position-space
transform at and below $k_J$ is the boundary-condition dependence already
noted there. Within the controlled regime $r\ll\lambda_J$, the leading
correction to Newton's law is parametrically
\begin{equation}
    \frac{\delta V_{\rm Newton}^{(\Pi)}}{V_{\rm Newton}}
    \;\sim\;
    (k_J r)^2
    \;=\;
    \frac{\epsilon_{\rm env}}{c_s^2},
    \label{eq:newton_pi_parametric}
\end{equation}
a $0\text{PN}\times\epsilon_{\rm env}/c_s^2$ effect. For typical
astrophysical environments ($\epsilon_{\rm env}/c_s^2\ll v^2$) this is
smaller than the 1PN worldline-vertex pieces in \eqref{eq:VEIH_full}; in
sufficiently dense, soft media---deep interiors of compact objects,
near-Jeans-unstable regions of dense disks---it can become competitive,
and a complete treatment requires the global Poisson--Jeans resummation.

\paragraph{$V^{(3)}_\Pi$ correction to $V_f$.}
The cubic vertex $V^{(3)}_\Pi$ analogously corrects the three-graviton
topology of Fig.~\ref{fig:EIH_diagrams}(f): sandwiching the static
$\Gamma_{TTT}^{0000,0000,0000}$ between three Newtonian potential lines
brings one factor of $G_N(\epsilon+p)$ via the medium correlator, so

$\delta V_f^{(\Pi)}$ sits at $1\text{PN}\times\epsilon_{\rm env}/c_s^4$:
in the static limit $\mathcal V_\psi\sim w\,c_s^2$ while each of the three
sound propagators gives $1/(wc_s^2)$, so
$\Gamma_{TTT}^{0000,0000,0000}|_{\rm static}\sim w/c_s^4$, one power of
$c_s^2$ more singular than $\Pi^{0000}_{\rm static}$. Its ratio to the
bulk-$\Pi$ correction to Newton is $v^2/c_s^2$, so that correction is the
leading bulk effect only for $v\ll c_s$.

\subsection{Gravitational Self-Energy}
\label{sec:app_DF}

The worldline self-energy isolates the tensor structure and the UV/IR
split of the in-medium correction; the full in-in derivation of the causal
self-force and its renormalization-group structure is deferred to the
companion paper~\cite{Modrekiladze:2026twz}. At leading order in the medium
expansion, the self-energy is obtained by inserting the dressed propagator
into the standard worldline two-point functional,
\begin{equation}
    \Sigma_{\rm med}
    = \frac{1}{2}\!\left(\frac{\kappa}{2}\right)^{\!2}
    \int\dd{\tau}\dd{\tau'}\,
    \mathcal{T}^{\mu\nu}_{\rm pp}(\tau)\,
    G_{\mu\nu\alpha\beta}\big(x(\tau),x(\tau')\big)\,
    \mathcal{T}^{\alpha\beta}_{\rm pp}(\tau').
\end{equation}

\begin{figure}[htbp]
    \centering
\tikzset{every picture/.style={line width=0.75pt}}
\begin{tikzpicture}[x=0.75pt,y=0.75pt,yscale=-1,xscale=1]
\draw    (200,150) -- (500,150) ;
\draw  [fill={rgb, 255:red, 0; green, 0; blue, 0 }  ,fill opacity=1 ] (248,150.5) .. controls (248,149.12) and (249.12,148) .. (250.5,148) .. controls (251.88,148) and (253,149.12) .. (253,150.5) .. controls (253,151.88) and (251.88,153) .. (250.5,153) .. controls (249.12,153) and (248,151.88) .. (248,150.5) -- cycle ;
\draw  [fill={rgb, 255:red, 0; green, 0; blue, 0 }  ,fill opacity=1 ] (448,150.5) .. controls (448,149.12) and (449.12,148) .. (450.5,148) .. controls (451.88,148) and (453,149.12) .. (453,150.5) .. controls (453,151.88) and (451.88,153) .. (450.5,153) .. controls (449.12,153) and (448,151.88) .. (448,150.5) -- cycle ;
\draw    (249,147) .. controls (249.32,144.57) and (250.69,143.45) .. (253.12,143.62) .. controls (255.52,143.85) and (256.89,142.77) .. (257.22,140.4) .. controls (257.57,138.04) and (258.93,137.02) .. (261.28,137.34) .. controls (263.6,137.71) and (264.94,136.73) .. (265.31,134.42) .. controls (265.7,132.12) and (267.03,131.2) .. (269.31,131.66) .. controls (271.56,132.16) and (272.88,131.29) .. (273.28,129.04) .. controls (273.7,126.79) and (275.66,125.56) .. (279.16,125.37) .. controls (281.3,126.04) and (282.6,125.29) .. (283.05,123.1) .. controls (283.52,120.92) and (284.81,120.21) .. (286.9,120.97) .. controls (288.95,121.77) and (290.22,121.1) .. (290.71,118.97) .. controls (292.51,116.22) and (294.4,115.31) .. (296.37,116.22) .. controls (298.3,117.17) and (299.55,116.61) .. (300.1,114.54) .. controls (300.69,112.47) and (302.53,111.71) .. (305.63,112.26) .. controls (307.43,113.35) and (308.64,112.89) .. (309.27,110.88) .. controls (309.92,108.88) and (311.72,108.27) .. (314.66,109.04) .. controls (316.33,110.27) and (317.52,109.91) .. (318.21,107.96) .. controls (320.12,105.7) and (321.87,105.23) .. (323.46,106.54) .. controls (326.13,107.63) and (327.85,107.23) .. (328.63,105.36) .. controls (329.45,103.49) and (331.14,103.18) .. (333.71,104.41) .. controls (335.08,105.88) and (336.74,105.63) .. (338.7,103.68) .. controls (339.61,101.89) and (341.25,101.72) .. (343.61,103.16) .. controls (344.8,104.74) and (346.41,104.64) .. (348.43,102.85) .. controls (350.5,101.12) and (352.08,101.08) .. (353.15,102.73) .. controls (355.18,104.41) and (356.73,104.43) .. (357.78,102.79) .. controls (359.91,101.22) and (361.92,101.34) .. (363.81,103.15) .. controls (364.64,104.9) and (366.11,105.06) .. (368.23,103.61) .. controls (370.4,102.23) and (372.31,102.52) .. (373.96,104.47) .. controls (374.59,106.28) and (375.99,106.55) .. (378.14,105.28) .. controls (380.35,104.06) and (382.16,104.49) .. (383.57,106.56) .. controls (384.02,108.4) and (385.33,108.77) .. (387.52,107.66) .. controls (389.75,106.61) and (391.46,107.16) .. (392.63,109.31) .. controls (393.71,111.46) and (395.35,112.07) .. (397.56,111.13) .. controls (399.79,110.24) and (401.37,110.89) .. (402.29,113.1) .. controls (403.12,115.31) and (404.64,116.01) .. (406.84,115.22) .. controls (409.07,114.47) and (410.52,115.21) .. (411.2,117.44) .. controls (411.79,119.66) and (413.17,120.43) .. (415.35,119.76) .. controls (417.55,119.12) and (418.87,119.91) .. (419.31,122.14) .. controls (419.67,124.35) and (421.22,125.36) .. (423.97,125.19) .. controls (426.12,124.64) and (427.28,125.46) .. (427.47,127.65) .. controls (428.12,130.21) and (429.48,131.23) .. (431.55,130.72) .. controls (434.13,130.65) and (435.38,131.66) .. (435.3,133.75) .. controls (435.6,136.2) and (436.95,137.37) .. (439.35,137.27) .. controls (441.72,137.18) and (442.9,138.29) .. (442.91,140.59) .. controls (442.79,142.82) and (443.96,144) .. (446.41,144.13) .. controls (448.77,144.21) and (449.82,145.36) .. (449.56,147.57) .. controls (449.43,150) and (450.5,151.32) .. (452.78,151.55) .. controls (451.18,152.49) and (450.48,151.73) .. (450.67,149.27) -- (449,147) ;
\draw (236,153.4) node [anchor=north west][inner sep=0.75pt]    {$\mathcal{T}_{pp}^{\ \mu \nu }( \tau )$};
\draw (437,153.4) node [anchor=north west][inner sep=0.75pt]    {$\mathcal{T}_{pp}^{\ \alpha \beta }( \tau ')$};
\draw (332,82.4) node [anchor=north west][inner sep=0.75pt]    {$G_{\mu \nu , \alpha \beta }$};
\end{tikzpicture}
    \caption{Self-energy. The compact object emits a graviton at $\tau$ and
    reabsorbs it at $\tau'$. The internal line is the dressed propagator
    $G_{\mu \nu ,\alpha \beta }$ of Eq.~\eqref{eq:Gfull}.}
    \label{fig:selfenergy}
\end{figure}
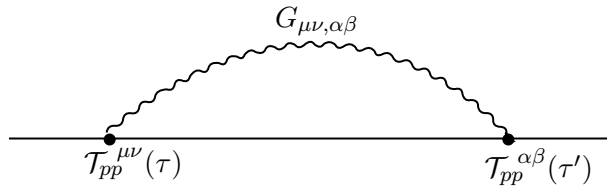

The topology is shown in Fig.~\ref{fig:selfenergy}.

Using the leading-order vacuum vertex, the stress tensor reduces to the bare
kinematic flux $\mathcal{T}^{\mu\nu}_{\rm pp}(\tau)=M v^\mu(\tau)v^\nu(\tau)
\delta^{(4)}(x-x(\tau))$, and the tensor algebra inside the loop simplifies
considerably. From Eq.~\eqref{eq:Gfull}, the leading medium-dependent piece
of the dressed propagator is
$\delta G_{\mu\nu\alpha\beta}=-D_{\mu\nu\rho\sigma}
(\kappa/2)^2\,\Pi^{\rho\sigma\gamma\delta}D_{\gamma\delta\alpha\beta}$,
with $D_{\mu\nu\rho\sigma}(k)=iP_{\mu\nu\rho\sigma}/(k^2+i\epsilon)$ the
vacuum de~Donder propagator.

Keeping only the leading $U^\mu U^\nu U^\alpha U^\beta$ tensor structure of the
sound pole, and suppressing the remaining $c_s^2$-dependent numerator terms, the
fluid correlator is approximated by
\begin{equation}
    \Pi_{\rho\sigma\gamma\delta}(k)
    \approx -\,(\epsilon+p)\,
    \frac{K^2}{(U\!\cdot\!k)^2-c_s^2 K^2+i\epsilon}\,
    U_\rho U_\sigma U_\gamma U_\delta,
\end{equation}
where $K^2=-k_\perp^2$.

The de~Donder contraction gives
\begin{equation}
    D_{\mu\nu\rho\sigma} U^\rho U^\sigma
    = \frac{i}{k^2+i\epsilon}\left( U_\mu U_\nu - \frac{1}{2}\eta_{\mu\nu} \right),
\end{equation}
and contracting in turn with the worldline vertex $Mv^\mu v^\nu$ (with the
proper-time normalization $v^2=1$) yields the single scalar factor
$\gamma^2-\tfrac12$, with $\gamma=v\cdot U$ the relative Lorentz factor.

The medium-induced effective action therefore reduces to the scalar integral
\begin{multline}
    \Sigma_{\rm med}
    = -\,\frac{M^2}{2}\left(\frac{\kappa}{2}\right)^{\!4}(\epsilon+p)
      \int d\tau d\tau'\,
      \left(\gamma_{\tau}^2 - \frac{1}{2}\right)
      \left(\gamma_{\tau'}^2 - \frac{1}{2}\right)
    \\
    \times
      \int \frac{d^4k}{(2\pi)^4}
      \frac{K^2\,e^{-ik\cdot(x(\tau)-x(\tau'))}}
           {\big((U\cdot k)^2-c_s^2K^2+i\epsilon\big)(k^2+i\epsilon)^2}.
\end{multline}
In the static limit the kernel collapses to $+\,w/c_s^2$: the medium
deepens the binding, the anti-screening of Sec.~\ref{sec:benchmarks}.

The physics separates cleanly according to the analytic structure of the
momentum integral. Away from the propagator poles, the real part is
conservative---energy temporarily stored in the fluid and returned---and its
short-distance UV divergences are absorbed by local counterterms,
renormalizing the worldline coefficients of the one-body matching; the
seagull \eqref{eq:seagull_fluid} contributes to $\Sigma_{\rm med}$ only a term
local on the worldline, absorbed by the same counterterms. On the
acoustic poles, the imaginary part describes radiation of on-shell phonons
into the bulk; varying it with respect to the Keldysh variable yields a
history-dependent dissipative drag---the fully relativistic generalization
of dynamical friction. For uniform motion the worldline support
$\omega=\mathbf k\cdot\mathbf v$ intersects the sound shell only for
supersonic relative motion: dynamical friction from an ideal fluid is
phonon \emph{Cherenkov radiation} into the Mach cone $\cos\theta=c_s/v$,
switching off for steady subsonic motion as in Landau's classic
sound-radiation argument. The on-pole evaluation for uniform supersonic
motion gives the fully relativistic drag, per unit proper time,
\begin{equation}
    \mathbf F_{\rm DF}
    = -\,4\pi G_N^2 M^2\, w\,
      \frac{(2\gamma^2-1)^2}{\gamma\,v^2}\,
      \ln\!\frac{k_{\rm UV}}{k_{\rm IR}}\;\hat{\mathbf v},
    \label{eq:DF_drag}
\end{equation}
whose Newtonian limit is Chandrasekhar's
law~\cite{Chandrasekhar:1943ys} with $\rho\to w$: the enthalpy that sets
the screening mass also sets the drag. The ultraviolet end of the
logarithm is no educated guess but a worldline counterterm fixed by
matching: for a black hole in dust, matching to exact geodesic
scattering places it at the Newtonian scale $R_c=G_NM/v^2$,
parametrically larger than the capture separatrix $b\simeq4G_NM/v$;
the infrared end is the Jeans-bounded size of the
medium~\cite{Modrekiladze:2026twz}.

\begin{figure}[htbp]
    \centering
\tikzset{every picture/.style={line width=0.75pt}}
\begin{tikzpicture}[x=0.75pt,y=0.75pt,yscale=-1,xscale=1]
\draw    (200,85) -- (500,85) ;
\draw    (200,185) -- (500,185) ;
\draw    (350,85) .. controls (351.67,86.67) and (351.67,88.33) .. (350,90) .. controls (348.33,91.67) and (348.33,93.33) .. (350,95) .. controls (351.67,96.67) and (351.67,98.33) .. (350,100) .. controls (348.33,101.67) and (348.33,103.33) .. (350,105) .. controls (351.67,106.67) and (351.67,108.33) .. (350,110) .. controls (348.33,111.67) and (348.33,113.33) .. (350,115) .. controls (351.67,116.67) and (351.67,118.33) .. (350,120) .. controls (348.33,121.67) and (348.33,123.33) .. (350,125) .. controls (351.67,126.67) and (351.67,128.33) .. (350,130) .. controls (348.33,131.67) and (348.33,133.33) .. (350,135) .. controls (351.67,136.67) and (351.67,138.33) .. (350,140) .. controls (348.33,141.67) and (348.33,143.33) .. (350,145) .. controls (351.67,146.67) and (351.67,148.33) .. (350,150) .. controls (348.33,151.67) and (348.33,153.33) .. (350,155) .. controls (351.67,156.67) and (351.67,158.33) .. (350,160) .. controls (348.33,161.67) and (348.33,163.33) .. (350,165) .. controls (351.67,166.67) and (351.67,168.33) .. (350,170) .. controls (348.33,171.67) and (348.33,173.33) .. (350,175) .. controls (351.67,176.67) and (351.67,178.33) .. (350,180) .. controls (348.33,181.67) and (348.33,183.33) .. (350,185) -- (350,185) ;
\draw    (350,135) .. controls (351.67,133.33) and (353.33,133.33) .. (355,135) .. controls (356.67,136.67) and (358.33,136.67) .. (360,135) .. controls (361.67,133.33) and (363.33,133.33) .. (365,135) .. controls (366.67,136.67) and (368.33,136.67) .. (370,135) .. controls (371.67,133.33) and (373.33,133.33) .. (375,135) .. controls (376.67,136.67) and (378.33,136.67) .. (380,135) .. controls (381.67,133.33) and (383.33,133.33) .. (385,135) .. controls (386.67,136.67) and (388.33,136.67) .. (390,135) .. controls (391.67,133.33) and (393.33,133.33) .. (395,135) .. controls (396.67,136.67) and (398.33,136.67) .. (400,135) .. controls (401.67,133.33) and (403.33,133.33) .. (405,135) .. controls (406.67,136.67) and (408.33,136.67) .. (410,135) .. controls (411.67,133.33) and (413.33,133.33) .. (415,135) .. controls (416.67,136.67) and (418.33,136.67) .. (420,135) .. controls (421.67,133.33) and (423.33,133.33) .. (425,135) .. controls (426.67,136.67) and (428.33,136.67) .. (430,135) -- (430,135) ;
\draw  [fill={rgb, 255:red, 0; green, 0; blue, 0 }  ,fill opacity=1 ] (347,134.5) .. controls (347,133.12) and (348.12,132) .. (349.5,132) .. controls (350.88,132) and (352,133.12) .. (352,134.5) .. controls (352,135.88) and (350.88,137) .. (349.5,137) .. controls (348.12,137) and (347,135.88) .. (347,134.5) -- cycle ;
\draw (338,63.4) node [anchor=north west][inner sep=0.75pt]    {$\mathcal{T}_{pp}^{\ \mu \nu }$};
\draw (338,188.4) node [anchor=north west][inner sep=0.75pt]    {$\mathcal{T}_{pp}^{\ \alpha \beta }$};
\draw (267,123.4) node [anchor=north west][inner sep=0.75pt]    {$V_{EH}^{( 3)} +V_{\Pi }^{( 3)}$};
\draw (435,123.4) node [anchor=north west][inner sep=0.75pt]    {$\Delta h_{ij}^{\rm med}$};
\end{tikzpicture}
    \caption{Tree-level Feynman diagram contributing to nonlinear gravitational-wave memory. Two radiation modes scatter in the bulk to produce a zero-frequency offset at null infinity. In a medium, the cubic interaction is modified by the fluid stress-energy three-point function.}
    \label{fig:memory_diagram}
\end{figure}
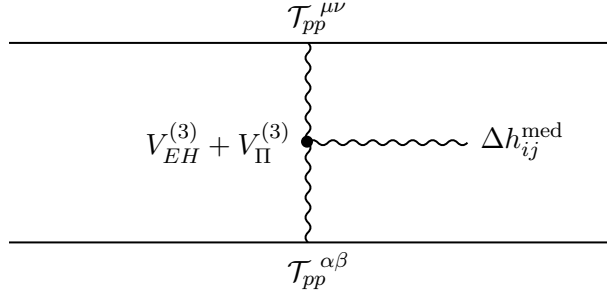

\subsection{Gravitational-Wave Memory in a Medium}
\label{sec:app_memory}

The Christodoulou nonlinear memory
effect~\cite{Christodoulou:1991cr,Blanchet:1992br} arises from the tree-level
topology in Fig.~\ref{fig:memory_diagram}: two hard radiation gravitons
emitted by the binary scatter in the bulk and generate a soft
($\omega\to0$) DC offset at null infinity---the soft graviton sourced by the
energy flux of the hard radiation. A medium alters this picture in exactly
three places: the masses that fix the hard source, the propagator that
carries the gravitons, and the cubic vertex itself. Relegating the
closed-time-path setup to App.~\ref{app:memory_kernels}, we work to first
order in the corresponding retarded medium corrections---the source
renormalization $\delta\mathcal T$, the propagator correction $\delta G_R$,
and the cubic-vertex correction $\delta V_R$---so that the in-medium memory
decomposes as
\begin{equation}
    \boxed{\;
    \Delta h_{ij}^{\rm med}
    =
    \Delta h_{ij}^{(F)}
    +
    \delta h_{ij}^{(G,{\rm soft})}
    +
    \delta h_{ij}^{(G,{\rm hard})}
    +
    \delta h_{ij}^{(V)}\,.\;
    }
    \label{eq:memory_split}
\end{equation}
Each piece is linear in a single medium correction; cross terms are
$\mathcal O(\epsilon_{\rm env}^2)$ and dropped. We abbreviate
$\delta h_{ij}^{(G)}\equiv\delta h_{ij}^{(G,{\rm soft})}
+\delta h_{ij}^{(G,{\rm hard})}$.

\paragraph{Vacuum memory kernel.}
To expose the structure of the in-medium corrections, first recall the
vacuum kernel. In the wave zone the hard radiation field is
$\bar h_{ab}^{TT}(u,r',\hat n')=A_{ab}(u,\hat n')/r'+\mathcal O(r'^{-2})$,
and the Isaacson stress tensor reduces to
\begin{equation}
    \tau_{\rm GW}^{\mu\nu}
    =
    \frac{1}{32\pi G_N}
    \big\langle
      \partial^\mu \bar h^{TT}_{ab}\,
      \partial^\nu \bar h^{TT}_{ab}
    \big\rangle
    =
    \frac{n'^\mu n'^\nu}{r'^2}\,
    \frac{dE_{\rm GW}}{du' d\Omega'},
    \qquad
    n'^\mu=(1,\hat n').
    \label{eq:memory_vacuum_flux}
\end{equation}
Evaluating the retarded integral at a distant observer
$x^\mu=(u+r,r\hat n)$ gives the standard Christodoulou formula,
\begin{equation}
    \boxed{\;
    \Delta h_{ij}^{\rm vac}(u,\hat n)
    =
    \frac{4G_N}{r}
    \left[
        \int_{-\infty}^{u}du'
        \int d\Omega'\,
        \frac{dE_{\rm GW}^{\rm vac}}{du' d\Omega'}\,
        \frac{n'_i n'_j}{1-\hat n\cdot \hat n'}
    \right]^{TT}\,.\;
    }
    \label{eq:memory_vacuum_kernel}
\end{equation}
The same geometric kernel appears in every piece below.

\paragraph{1. Source renormalization.}
The source piece is the vacuum kernel \eqref{eq:memory_vacuum_kernel}
acting on the hard flux corrected at first order in the density: one of
the two factors of $\dddot Q^{ij}$ in the vacuum quadrupole flux is
replaced by $\dddot Q_{\rm med}^{ij}-\dddot Q_{\rm vac}^{ij}$, and the two
contributions are summed. Using
$M_{{\rm rad},A}=-(F_A-F_A')=M_A+\tfrac{1}{2}\rho_A V_{pA}$ from
Eq.~\eqref{eq:F1},
\begin{equation}
    \Delta h_{ij}^{(F)}(u,\hat n)
    = \Delta h_{ij}^{\rm vac}(u,\hat n)
      \Big|_{M_A\,\to\,M_A+\frac12\rho_A V_{pA}}
      -\Delta h_{ij}^{\rm vac}(u,\hat n)
      +\mathcal O(\rho^2),
    \label{eq:memory_F}
\end{equation}
with the replacement performed in the full binary quadrupole before
squaring, at fixed orbital trajectories, inside the hard flux under the retarded
memory integral; it survives in the fluid rest frame and is the leading
isotropic environmental renormalization of the memory amplitude.

\paragraph{2. Dressed propagator and medium vertex.}
Both vanish for a homogeneous perfect fluid at this order. By the Ward
identity \eqref{eq:ward} a gauge shift changes the induced medium stress only
by terms proportional to $\bar T$, so one may evaluate in the fluid rest
frame, where a perfect fluid has no anisotropic stress at linear order:
$t_s^{\mu\nu}$ has no transverse-traceless part, the vortex sector
contributes only to $\Pi^{0i0j}$, and the seagull adds only the isotropic
contact term $-2p$, cancelled by the curvature it accompanies. Hence
$\delta h^{(G)}=\delta h^{(V)}=0$ in every frame. Contracting the sound-pole
kernel alone with the lab-frame projector instead produces the flow structure
$U^T_{\langle i}U^T_{j\rangle}$; that is the response of a non-transverse
kernel to the pure-gauge part of the de~Donder field, and it disappears once
the seagull is included. A uniformly drifting homogeneous fluid is a fluid at
rest seen from another frame.

\paragraph{3. Rigid media and the astrophysical weathervane.}
A gap $m^2$ in the transverse-traceless response,
$\omega^2=|\mathbf k|^2+m^2$, screens the zero-frequency offset as $e^{-mL}$
over a path $L$, with $1/m$ the screening length. A medium that screens the
two polarizations over different lengths therefore writes its orientation
onto the permanent strain, and that requires rigidity. For a general
shear-modulus tensor the gap is a matrix on the two polarizations,
$m^2_{(\lambda\lambda')}$ of Eq.~\eqref{eq:solid_mass}, whose eigenvectors
are the medium's elastic axes projected on the sky. Write $m_a,m_b$ for the
two inverse screening lengths and $\chi$ for the angle between those axes and
the binary's memory axes. Propagating the soft field gives
$\Delta h^{\rm mem}_{(\lambda)}=[e^{-\sqrt{m^2}L}]_{(\lambda\lambda')}
\Delta h^{\rm vac}_{(\lambda')}$, whose off-diagonal element is
$\tfrac12(m_a-m_b)L\sin2\chi$ at first order, so a purely ``$+$'' vacuum
memory acquires
\begin{equation}
    \boxed{\;\frac{\Delta h_\times}{\Delta h_+}
    = -\,\tfrac12\,(m_a-m_b)\,L\,\sin2\chi\,.\;}
    \label{eq:memory_weathervane}
\end{equation}
a cross-polarized DC offset forbidden in vacuum and not degenerate with any
vacuum parameter. The screening length $1/\sqrt{16\pi G_N\mu}$ is about
$5\times10^3$ km for a neutron-star crust shear modulus, so
$m_aL,\,m_bL\ll1$ and the effect is a proof of principle: a memory that
records its environment measures rigidity and the orientation of the elastic
axes, not a bulk flow.

\paragraph{Explicit waveform for a circular binary.}
To turn the weathervane into a usable strain, specialize to a
quasi-circular binary of total mass $M$, symmetric mass ratio $\eta$,
inclination $\iota$, distance $D$, and post-Newtonian parameter
$x=(G_N M\omega/c^3)^{2/3}$. Its dominant nonlinear memory is purely
``$+$'' polarized in the source frame and grows monotonically through the
inspiral~\cite{Favata:2010zu},
\begin{equation}
    h_+^{\rm mem}(\iota)\;\propto\;\frac{G_N\eta M}{c^2 D}\,x\,
    \sin^2\iota\,(17+\cos^2\iota),
    \qquad h_\times^{\rm mem}=0,
    \label{eq:memory_circular_vac}
\end{equation}
the standard result, which the source piece $\Delta h^{(F)}$ reproduces
with the renormalized radiative masses
$M_{{\rm rad},A}=M_A-\tfrac52\rho_A V_{pA}$.

In a perfect fluid the in-medium memory is therefore the Christodoulou memory
with renormalized radiative masses, an isotropic correction degenerate with
the vacuum masses; in a rigid medium
Eq.~\eqref{eq:memory_weathervane} adds the cross-polarized offset, the one
environmental piece of the memory that no vacuum parameter can mimic.

\section{Applications II: Topologies with Zero Vacuum Limits}
\label{sec:applications_zero}

We now turn to effects from topologies forced to vanish in vacuum GR by
symmetry or kinematics, activated by the in-medium Feynman rules.

\subsection{Tidal Love Numbers of Black Holes}
\label{sec:app_love}

The relevant operator is the worldline tidal coupling
$\alpha_E\int E_{\mu\nu}E^{\mu\nu}\,d\tau$, whose coefficient $\alpha_E$ is the
electric-type Love number. In vacuum it vanishes identically in the static
limit: a hidden ``ladder'' symmetry truncates the regular static
perturbations of Schwarzschild to a polynomial, forcing
$\alpha_E(\omega=0)=0$~\cite{Hui:2021vcv,Hui:2022vbh}.

A medium spoils this. Replacing the internal graviton lines by the dressed
propagator $G$ feeds the sound pole of $\Pi$ into the perturbation equations,
whose explicit frequency dependence $\omega^2-c_s^2K^2$ is incompatible with the
ladder recursion. The Love number no longer vanishes. 
Its size follows from three inputs. The hole supplies the only length,
so dimensions fix the factor $(G_NM)^5/G_N$. The coupling
$h_{\mu\nu}\delta T^{\mu\nu}$ of Eq.~\eqref{eq:Sint} supplies one power of
the medium enthalpy, and hence the dimensionless $\epsilon_{\rm fluid}$. The
frequency dependence is the medium's longitudinal response
$\Pi^{0000}(\omega,K)$ normalized to its static value
$(\epsilon+p)/c_s^2$. Together,
\begin{equation}
    \alpha_E(\omega) \sim \frac{1}{G_N}(G_N M)^5\,\epsilon_{\rm fluid}\,
    \frac{c_s^2K^2}{c_s^2K^2 - \omega^2},
    \qquad
    \epsilon_{\rm fluid}\equiv\frac{(\epsilon+p)R_s^3}{M},
    \label{eq:alphaE}
\end{equation}
with $R_s=2G_N M$ the Schwarzschild radius.
Here $\omega$ is the frequency of the applied tide, twice the orbital
frequency for a circular binary, and $K$ is the wavenumber that tide forces
in the medium, set by the orbital separation, $K\sim1/r$. The response is
resonant at $\omega=c_sK$, where the forcing matches the medium's sound mode.
With these identifications the condition reads $v\simeq c_s/2$ for the
orbital velocity: the resonance falls where the orbit turns transonic, at a
gravitational-wave frequency $c_s^3/8\pi G_NM$.
The response is induced entirely by the coupling of the hole's tidal field
to the surrounding fluid, not by any internal dynamics---the environment
breaks the symmetry that protected it. We give the scaling and mechanism
here; the full matching of $\alpha_E$ is carried out in the companion
analysis~\cite{Modrekiladze:2025love}.


\subsection{Graviton Splitting in a Medium: $h\to hh$}
\label{sec:graviton_splitting}

In vacuum $h\to hh$ is forbidden: three exactly massless on-shell legs are
collinear, and the phase space collapses. Integrating out the medium leaves a
graviton propagator with a second branch, the longitudinal Jeans--sound pole of
Sec.~\ref{sec:benchmarks}, and splitting into that branch is open. Two channels
follow, and we compute both. Splitting into three transverse legs stays closed.

\paragraph{No dressing of the transverse sector at linear response.}
On transverse-traceless legs a homogeneous perfect fluid contributes nothing at
first order in $\epsilon_{\rm env}$. In the fluid rest frame $t_s^{\mu\nu}$ has
no TT component, the vortex sector contributes only to $\Pi^{0i0j}$, and the
seagull adds only the isotropic contact term $-2p$, the term a cosmological
constant would produce, cancelled by the curvature of the background sourced by
$\bar T$. The dressed TT shell therefore stays at $k^2=0$. Since the physical
helicity states are frame independent up to gauge and \eqref{eq:ward} controls
the gauge variation, the same holds for a moving medium: no effective mass, no
birefringence, and no opening of $h\to hh$. Evaluating the non-transverse
kernel on lab-frame TT polarizations would instead give a shift
$\propto(\epsilon+p)(1-c_s^2)|U_iU_j\epsilon^{TT\,ij}|^2$, which is exactly the
gauge variation $2(U\!\cdot\!k)(U\!\cdot\!\xi)$ of $U\!\cdot\!\epsilon\!\cdot\!U$
and is removed by the seagull. This agrees with Flauger and
Weinberg~\cite{Flauger:2017gtb}: a pressureless perfect fluid leaves the wave
undispersed, and the leading effect is the isotropic velocity-dispersion shift,
for which the triangle inequality fails so that channel stays closed too.

\paragraph{First channel: splitting into two longitudinal gravitons.}
Linear response is not all of $\Pi$. The anisotropic stress has a two-quantum
continuum whose support contains $\omega=|\mathbf k|$ and
collapses to the collinear vacuum point as $c_s\to1$. That is a dressing of
the TT propagator, and the dressed shell acquires a width
$\Gamma=(8\pi G_N/\omega)\operatorname{Im}[\epsilon^{TT*}\!\cdot\Pi\cdot
\epsilon^{TT}]_{|\mathbf k|=\omega}$. Evaluating the continuum from the fluid
EFT,
\begin{equation}
    \boxed{\;
    \operatorname{Im}\bigl[\epsilon^{TT*}\!\cdot\Pi\cdot\epsilon^{TT}\bigr]_{|\mathbf k|=\omega}
    =\frac{\omega^4}{960\pi}\frac{(1-c_s^2)^2}{c_s^3},
    \qquad
    \Gamma=\frac{G_N\omega^3}{120}\frac{(1-c_s^2)^2}{c_s^3}\,.\;}
    \label{eq:hpipi_rate}
\end{equation}
The width is the same for both helicities, is independent of the density, and
vanishes as $(1-c_s^2)^2$ so a stiff fluid is exactly transparent. The density
cancels because the coupling carries one power of $w$ while each canonically
normalized mode carries $w^{-1/2}$. It re-enters only through the validity of
the fluid description, which needs the final-state momenta $q\sim\omega/c_s$ to
lie below the inverse mean free path: the rate is unchanged as the medium
thins, but the frequency window in which it applies closes. For a moving
medium it transforms as a massless width, picking up
$[\gamma(1-\mathbf v_{\rm med}\!\cdot\!\hat n)]^4$, an anisotropy that is the
Doppler factor of an isotropic rest-frame rate rather than a genuine
anisotropy of the medium.

\paragraph{Second channel: splitting into a transverse and a longitudinal graviton.}
This is the channel that uses the medium as the self-force does,
$h(k)\to h(k_2)+h^*(q)$ through the vacuum three-graviton vertex with the third
line dressed. This contribution is the self-energy diagram of Sec.~\ref{sec:app_DF} with the
worldline replaced by $V^{(3)}_{\rm EH}$. The fluid also contributes the contact
interaction
\begin{equation*}
    \mathcal L_{hh\pi}^{\rm TT}
    =-\frac{\kappa^2w c_s^2}{4}\,(\partial_i\pi^i)\,h_{jk}^{\rm TT}h_{jk}^{\rm TT},
\end{equation*}
where $\pi^i$ is the displacement of App.~\ref{app:phonon_vertex}.
The rate includes both contributions and their interference. The kinematics
$\omega-|\mathbf k_2|=c_s|\mathbf k-\mathbf k_2|$ admit solutions, closing
only as $c_s\to1$. The event rate is infrared divergent linearly in the
absorbed energy;
the observable is the energy transferred to the medium, which converges
logarithmically. Per graviton of energy $\omega$, in the fluid rest frame,
\begin{equation}
    \boxed{\;\Gamma_E\equiv-\frac{d\ln\omega}{dt}
    =16\pi G_N^2\,w\,\omega\Bigl[\ln\frac{\omega}{E_{\min}}+f(c_s)\Bigr],\;}
    \label{eq:graviton_DF}
\end{equation}
with $E_{\min}$ the infrared cutoff of the sound pole and $f(c_s)$ a slowly
varying offset known in closed form, tending to $\ln2c_s-\tfrac{11}{12}$ for
$c_s\ll1$ and to $-\tfrac32$ at $c_s\to1$. The coefficient of the logarithm
is exact and is the massless limit of Eq.~\eqref{eq:DF_drag}: taking
$\gamma\to\infty$ at fixed $M\gamma\to\omega$ gives
$16\pi G_N^2\omega^2w\ln$, i.e.\ $\omega\Gamma_E$. A graviton suffers
dynamical friction like any other energy-carrying probe.

\paragraph{Kinetic media: no birefringence, and Flauger--Weinberg as the missing seagull.}
A collisionless medium is a superposition of cold streams, and on
common-frame TT polarizations the exact seagull-completed dust kernel is
$4\rho_s(U_s\!\cdot\!\epsilon_{(\lambda)})\!\cdot\!(U_s\!\cdot\!\epsilon_{(\lambda')})
=-2\rho_s\gamma_s^2|\mathbf v_s^T|^2\delta_{\lambda\lambda'}$, with $\gamma_s$
and $\mathbf v_s^T$ common-frame quantities. It is proportional to the identity
on helicity space, so \emph{no} velocity distribution splits the two helicities
at tree level. Summing over an isotropic distribution gives $-4p_{\rm kin}$;
the self-consistent background removes what a perfect fluid with the same
$\bar T$ would give, namely its seagull $-2p_{\rm kin}$, and a collisionless
medium never had that seagull. The difference is
$m^2=16\pi G_Np_{\rm kin}=\tfrac{32\pi G_N}{3}\mathcal E$, the
Flauger--Weinberg mass, obtained here as the seagull a kinetic medium lacks.

\paragraph{Elastic media: rigidity gives birefringence and the memory weathervane.}
A helicity-distinguishing response needs dynamical anisotropic stress at linear
order, i.e.\ rigidity. In a solid the TT perturbation is a genuine shear
strain, so the seagull carries the shear modulus: for the relativistic solid
EFT~\cite{Dubovsky:2005xd} its TT block is $-2p-2\mu$. In an isotropic solid a TT graviton cannot
mix with a single transverse phonon, so there is no exchange term on TT,
the $-2p$ is background-cancelled, and the remainder is a physical,
unambiguous gap,
\begin{equation}
    \boxed{\;m^2_{\rm solid}=16\pi G_N\mu,\qquad
    m^2_{(\lambda\lambda')}(\mathbf k=0)=8\pi G_N\,\epsilon^{(\lambda)}_{ij}\mathcal M_{ijkl}
    \epsilon^{(\lambda')}_{kl}\;}
    \label{eq:solid_mass}
\end{equation}
since $\mu$ is a response coefficient, not a component of $\bar T$, so no
background absorbs it. This is a gap the medium carries, not a vacuum graviton
mass: it vanishes with $\mu$ outside the medium, the propagating modes are
still the two transverse-traceless ones, and the medium's Goldstones keep
diffeomorphism invariance intact, exactly as a plasma frequency does for a
photon.
For an anisotropic solid at finite momentum, phonon exchange contributes as
well. Define $b_{i\lambda}=k_j\mathcal M_{ijkl}\epsilon^{(\lambda)}_{kl}$ and
$D_{i\ell}=k_j\mathcal M_{ij\ell m}k_m$. For unit-normalized TT tensors,
\begin{equation}
 \begin{split}
 \Sigma^{\rm shear}_{\lambda\lambda'}(\omega,\mathbf k)
 =8\pi G_N\Bigl[&
 \epsilon^{(\lambda)*}_{ij}\mathcal M_{ijkl}\epsilon^{(\lambda')}_{kl}\\
 &+b^*_{i\lambda}\bigl[(\epsilon+p)(\omega+i0)^2\mathbf1-D\bigr]^{-1}_{i\ell}
 b_{\ell\lambda'}\Bigr].
 \end{split}
 \label{eq:elastic_phonon_exchange}
\end{equation}
 An anisotropic crystal is birefringent, and because the gap survives
at zero momentum its effect grows towards low frequency and is largest on the
memory, where it gives the weathervane of
Eq.~\eqref{eq:memory_weathervane}. Magnetized plasmas and parity-odd media
belong to the same class.

\section{Observational Prospects}
\label{sec:observations}

Every medium insertion carries one factor of the ambient density, so to
leading order  most of the effects derived above scale linearly
with $\rho$ (or $\epsilon+p$)---a single master tuning parameter.
Two exceptions: the width \eqref{eq:hpipi_rate} is independent of the
density, and the rigidity-controlled effects scale with the shear modulus
$\mu$ rather than with $\rho$. Representative
environments span over thirty decades in density
(Table~\ref{tab:predictions}), so the same formula moves from comfortably
observable to proof-of-principle as one changes environment.

\paragraph{Conservative and dissipative dephasing.}
The 1PN potential of Sec.~\ref{sec:app_EIH} and the drag of
Sec.~\ref{sec:app_DF} imprint themselves on the inspiral phase at
\emph{negative} post-Newtonian orders relative to the leading vacuum
flux---dynamical friction at $-5.5$PN, accretion and gravitational pull at
$-3$ to $-2.5$PN---so they dominate at low frequency and are separable from
the vacuum series. Bayesian analyses can distinguish a dark-matter spike, an
accretion disk, and a boson cloud from one another and from vacuum mass and
spin parameters~\cite{Barausse:2014tra,Cardoso:2019rou,Cole:2022yzw}, with
the dark-matter spike slope measurable to a few percent even after
energy-conserving halo feedback is
included~\cite{Eda:2013gg,Kavanagh:2020cfn,Coogan:2021uqv}. Below the
current LIGO--Virgo band for any plausible density, these effects are
squarely next-generation targets: the Einstein Telescope and Cosmic
Explorer, with their greatly extended low-frequency reach and event rates,
should register environmental dephasing in hundreds of stellar-mass
signals at the population level~\cite{Zwick:2025env}, while the
lowest-frequency effects target LISA and deci-hertz observatories.

\paragraph{Tidal signatures and the sound resonance.}
Because the static tidal Love numbers of a black hole vanish
identically~\cite{Binnington:2009bb,Damour:2009vw}, any nonzero value is a
clean diagnostic of new physics or environment. The medium-induced Love
number of Sec.~\ref{sec:app_love} is sharply distinguished from a smooth
background contribution by its resonance at $\omega=c_sK$---for a coherent
scalar cloud, at the gravitational-atom frequency $\omega=\mathbf k^2/2m$
following from Eq.~\eqref{eq:cs_scalar}; such a cloud can dominate the
environmental tidal response~\cite{DeLuca:2021ite,Arana:2024kaz}. With the
Einstein Telescope sharpening stellar-mass tidal measurements and LISA
forecast to tighten supermassive Love numbers by orders of
magnitude~\cite{Cardoso:2017cfl}, a frequency-localized resonance is an
especially clean target across the mass spectrum.

\paragraph{Propagation: opacity.}
The gravitational opacity must be set against a sharp benchmark: in four
dimensions the optical depth of any astrophysical medium to gravitons is less
than unity~\cite{Flauger:2019cam}, and any speed shift is bounded by
$|c_{\rm gw}-c|/c\lesssim10^{-15}$~\cite{Flauger:2017gtb}. The width
\eqref{eq:hpipi_rate} respects both: it is genuine but minute,
$\Gamma/\omega\sim(\omega/M_{\rm pl})^2$, and it leaves the propagation speed
untouched. The novelty is structure rather than size: the channel exists for
every medium, its rate is independent of the density, it closes
exactly for a stiff fluid, and its anisotropy is the Doppler factor of the
flow. Birefringence, by contrast, is a signature of rigidity rather than of
density or motion.

\paragraph{Memory.}
Nonlinear memory is a permanent strain offset of order ten percent of the peak
oscillatory strain~\cite{Favata:2010zu}, marginal for individual ground-based
events but accumulating by stacking~\cite{Lasky:2016knh}, and reaching
signal-to-noise of order ten for a single supermassive merger with LISA. In a
perfect fluid the medium enters only through the renormalized radiative masses,
degenerate with the vacuum masses. A memory pattern that records its
environment requires rigidity: an elastic medium screens the two polarizations
differently and populates the cross-polarized DC channel,
Eq.~\eqref{eq:memory_weathervane}, an \emph{astrophysical weathervane} whose
axes are the medium's elastic axes. The amplitude lies far below reach for any
plausible rigidity, so this is a proof of principle rather than a target.

\paragraph{Magnitudes across environments.}
Table~\ref{tab:predictions} collects representative order-of-magnitude
estimates across environments.
In summary, the dephasing and tidal sound resonance are plausibly
observable with the Einstein Telescope, LISA, and deci-hertz detectors in
dense environments; the
opacity is a proof of principle, and rigid media are where birefringence and a
medium-recording memory live.

\begin{table}[t]
\centering
\small
\setlength{\tabcolsep}{4pt}\footnotesize
\begin{tabular}{@{}l l l l@{}}
\toprule
Environment & $\rho\,[\mathrm{g\,cm^{-3}}]$ & Leading in-medium effect & Observability \\
\midrule
AGN thin disk & $10^{-10}$--$10^{-6}$ & gas/accretion dephasing, $\sim10^{1}$--$10^{2}$ rad & ET (population); LISA EMRIs \\
ADAF/RIAF & $10^{-18}$--$10^{-13}$ & dynamical-friction dephasing & LISA (dense, favorable) \\
DM spike & $\sim10^{-20}$ & DF dephasing, $\sim10^{3}$--$10^{7}$ cycles$^{\dagger}$ & LISA ``dark dress'' \\
Boson cloud & $\sim1$--$10^{10}$$^{\ddagger}$ & Love resonance at $\omega=\mathbf k^2/2m$ & ET tidal; LISA EMRIs \\
Local DM & $\sim7\times10^{-25}$ & GW refraction, $\delta c/c\lesssim10^{-20}$ & unobservable \\
\bottomrule
\end{tabular}
\caption{Order-of-magnitude reach across representative environments; 
the listed effects scale linearly with $\rho$. $^{\dagger}$Static estimate;
energy-conserving halo feedback reduces it by $\sim$two orders of magnitude
but leaves a LISA-detectable
residual~\cite{Eda:2013gg,Kavanagh:2020cfn,Coogan:2021uqv}.
$^{\ddagger}$Rough, mass-dependent estimates. }
\label{tab:predictions}
\end{table}

\section{Conclusion}
\label{sec:conclusions}

We have formulated an effective field theory of gravity in relativistic
media with a simple organizing principle: once the medium is integrated
out, the diagram topologies are those of vacuum, while the propagators and
vertices are dressed by the response of the environment. Environmental
effects thus become systematic deformations of the gravitational Feynman
rules rather than a disconnected collection of special force laws.

For vacuum topologies that survive in a medium, we derived the full 1PN
Einstein--Infeld--Hoffmann potential in a relativistic fluid,
fluid-mediated three-body forces, the 1PN Stokes drag, the gravitational
self-energy, and the in-medium Christodoulou memory, 
which for an ideal fluid reduces to a renormalization of the radiative
masses and, in a rigid medium, acquires a cross-polarized offset aligned with
the elastic axes---an \emph{astrophysical weathervane}. For topologies forced to vanish in
vacuum, the sound pole breaks the ladder symmetry protecting black-hole
Love numbers, producing a resonant, running tidal response, and 
the medium opens the channel $h\to hh$ into the longitudinal branch
of the dressed propagator: 
it gives the transverse graviton the closed-form width
\eqref{eq:hpipi_rate}, a \emph{gravitational opacity} set by the sound speed, and
makes a graviton suffer dynamical friction, Eq.~\eqref{eq:graviton_DF}.

Beyond a single fluid, the construction unifies a catalogue of media
through one object: the static screening response that fixes the Jeans
scale. The same replacement rule carries the sound pole into a Landau cut
or the de~Broglie dispersion; the gravitational-atom spectrum emerges as
the in-medium Love resonance, and the fluid, kinetic, and quantum Jeans
dispersions are recovered with no free parameter. The
environment also induces a logarithmic running of the drag
coefficient~\cite{Modrekiladze:2026twz}, and the framework extends order by order to viscous
fluids~\cite{Modrekiladze:2024htc}, superfluids, anisotropic media, and
spinning compact objects. The dephasing and tidal resonances lie within
reach of the Einstein Telescope and LISA;  the opacity is a proof of
principle, and rigid media are where birefringence and a medium-recording
memory live.

If gravitational-wave astronomy is to use environmental effects as signal
rather than nuisance, it needs a language as systematic as vacuum gravity itself. That
language exists: gravity in media is vacuum gravity with updated
propagators and vertices.
\section*{Acknowledgments}
I thank Ira Rothstein for many fruitful discussions, his input and for collaboration in the early stages of this project. I also thank Zvi Bern,
Mateja Bo\v{s}kovi\'{c}, Vitor Cardoso, Rafael Porto and Jordan Wilson-Gerow
for useful comments. The work was supported by the Lightcone Foundation and by the Deutsche Forschungsgemeinschaft (DFG) under Germany's Excellence Strategy, EXC 2121 ``Quantum Universe'' (390833306).
\appendix
\section{The Three-Phonon Vertex from the Fluid EFT}
\label{app:phonon_vertex}
This appendix derives the amputated longitudinal three-phonon vertex
$\mathcal V_\psi(k_1,k_2,k_3)$ used in Sec.~\ref{sec:feynman}. The
starting point is the cubic perfect-fluid Lagrangian
of~\cite{Endlich:2010hf}; we project onto three external longitudinal
phonons, fix the normalization relating the phonon field $\pi^i$ to the
sound-mode variable $\psi$ of Eq.~\eqref{eq:psi_prop}, and express the
result in the kinematic variables $(E_i,K_i^2,Q_{ij})$ of the main text.

The perfect-fluid EFT is built from three scalar fields
$\phi^I(x)$, $I=1,2,3$, with internal symmetries enforcing a
fluid equation of state~\cite{Son:2005tj,Dubovsky:2005xd,Endlich:2010hf}.
Writing $\mathcal L=-w_0\,f(\sqrt B)$ with
$B\equiv\det(\partial_\mu\phi^I\partial^\mu\phi^J)$,
$w_0=(\epsilon+p)|_{B=1}$ the background enthalpy density, and $f$
normalized so that $f'(1)=1$, the speed of sound is $c_s^2=f''(1)$. Expanding about the ground state
$\phi^I = x^I + \pi^I$ and keeping terms up to cubic order yields
the second line of Eq.~(17) of~\cite{Endlich:2010hf},
\begin{equation}
    \mathcal L_3
    = w_0\!\left[
        \tfrac{1}{2}c_s^2\,[\partial\pi][\partial\pi^2]
        -\tfrac{1}{6}(3c_s^2 + f_3)[\partial\pi]^3
        +\tfrac{1}{2}(1+c_s^2)[\partial\pi]\,\dot{\vec\pi}^2
        - \dot{\vec\pi}\cdot\partial\pi\cdot\dot{\vec\pi}
    \right],
    \label{eq:app_L3}
\end{equation}
where $(\partial\pi)^i{}_j\equiv \partial_j\pi^i$, $[X]\equiv\mathrm{tr}X$,
$[\partial\pi]=\partial_i\pi^i$,
$[\partial\pi^2]=\partial_i\pi^j\partial_j\pi^i$,
$\dot{\vec\pi}^2=\dot\pi^i\dot\pi^i$,
$\dot{\vec\pi}\cdot\partial\pi\cdot\dot{\vec\pi}
 =\dot\pi^i(\partial_i\pi^j)\dot\pi^j$, and
$f_3\equiv f'''(1)$; the equation of state rewrites this as the
main-text form
$f_3=w\,\partial c_s^2/\partial\epsilon+c_s^2(c_s^2-1)$.
We work in the perfect-fluid limit $c_T\to 0$, in which the transverse
deformation of~\cite{Endlich:2010hf} decouples from tree-level
longitudinal amplitudes.

The quadratic action for the longitudinal mode, with
$\pi^i = \partial^i \psi_\pi / |\partial|$ and $\psi_\pi$ a
canonically-normalized longitudinal scalar, yields the Feynman propagator
\begin{equation}
    \langle \psi_\pi(-k)\,\psi_\pi(k)\rangle
    = \frac{i}{w_0\big[(U\cdot k)^2 - c_s^2 K^2 + i\epsilon\big]}.
\end{equation}
The main text~\eqref{eq:psi_prop} instead uses a variable $\psi$ with
propagator carrying a $K^2$ in the numerator; the two are related by
\begin{equation}
    \psi(k) = K\,\psi_\pi(k),
    \qquad
    \text{i.e.}\quad
    \pi^i(k) = \frac{k^i}{K}\,\psi_\pi(k) = \frac{k^i}{K^2}\,\psi(k)
    \quad\text{(longitudinal projection)},
    \label{eq:app_pi_psi}
\end{equation}
the identification connecting the Lagrangian \eqref{eq:app_L3} to the
main-text form factor $t_s^{\mu\nu}(k)$ and propagator \eqref{eq:psi_prop}.

For three external longitudinal phonons with incoming momenta
$k_1, k_2, k_3$ satisfying $k_1+k_2+k_3=0$, the polarization
prescription~\eqref{eq:app_pi_psi} gives, for each leg,
$\partial_i\pi_a^j\to i\,k_a^i k_a^j/K_a$ and
$\dot\pi_a^i\to -i\,E_a k_a^i/K_a$
in the fluid rest frame, with $E_a\equiv U\cdot k_a$ and
$K_a \equiv |\vec{k}_a|$. Introducing
$Q_{ab}\equiv k_{a,\perp}\cdot k_{b,\perp}$, which in the rest
frame equals $-\vec{k}_a\cdot\vec{k}_b$, and summing all $3!$ Wick
contractions of the three external legs onto the three field slots (for
the single-trace factors leg $a$ sits in the $[\partial\pi]$ slot; for
the last operator leg $b$ sits in the middle $\partial\pi$ slot), the
four operators of \eqref{eq:app_L3} contribute to the amputated cubic
vertex as
\begin{align}
    [\partial\pi]^3 &\;\longrightarrow\; -6\,i\,K_1K_2K_3,
    &
    [\partial\pi][\partial\pi^2] &\;\longrightarrow\;
      -2i\sum_{\rm cyc}\frac{K_a\,Q_{bc}^2}{K_bK_c},
    \notag\\
    [\partial\pi]\,\dot{\vec\pi}^2 &\;\longrightarrow\;
      2i\sum_{\rm cyc}\frac{K_a\,E_bE_c\,Q_{bc}}{K_bK_c},
    &
    \dot{\vec\pi}\cdot\partial\pi\cdot\dot{\vec\pi} &\;\longrightarrow\;
      -2i\sum_{\rm cyc}\frac{E_aE_c\,Q_{ab}Q_{bc}}{K_aK_bK_c}.
\end{align}

Assembling the four contributions above with the coefficients in
\eqref{eq:app_L3} into the $\psi_\pi$-basis vertex
$\mathcal V_{\psi_\pi}$, factoring out the overall $i$ of the Feynman
rule, converting to the main-text $\psi$ basis via
\eqref{eq:app_pi_psi}---each external leg carries a factor $1/K_a$, so
$\mathcal V_\psi=\mathcal V_{\psi_\pi}/(K_1K_2K_3)$---and using the
identity $K_a^2/(K_1^2K_2^2K_3^2)=1/(K_b^2K_c^2)$, we finally obtain
(with $w=w_0$ on the background)
\begin{equation}
    \mathcal V_\psi
    = w\!\left[
        (1+c_s^2)\sum_{\rm cyc}
          \frac{E_a E_b\,Q_{ab}}{K_a^2 K_b^2}
      + 2 \sum_{\rm cyc}
          \frac{E_a E_c\,Q_{ab}Q_{bc}}{K_1^2 K_2^2 K_3^2}
      - c_s^2 \sum_{\rm cyc}
          \frac{Q_{ab}^2}{K_a^2 K_b^2}
      + (3c_s^2 + f_3)
    \right].
\end{equation}
As used in 
Sec.~\ref{sec:graviton_splitting}, $\mathcal V_\psi$ itself has no
\emph{sound} poles: all
nonanalyticity of $\Gamma_{TTT}^{\rm exch}$
\eqref{eq:GammaTTT_exchange} resides in the external sound propagators
$G_F^\psi(k_a)$.

\section{In-Medium Memory Kernels}
\label{app:memory_kernels}
This appendix records the closed-time-path setup underlying the in-medium
memory of Sec.~\ref{sec:app_memory}. For causal observables the relevant
objects are the retarded in-medium kernels of Sec.~\ref{sec:feynman}:
suppressing the closed-time-path indices, the retarded effective action is
\begin{align}
    S_{\rm eff}^R[h]
    =\;&
    \frac{1}{2}\int d^4x\,d^4y\;
    h_{\mu\nu}(x)\,
    \mathcal K_R^{\mu\nu}{}_{\alpha\beta}(x,y)\,
    h^{\alpha\beta}(y)
    \notag\\
    &+
    \frac{1}{3!}\int d^4x\,d^4y\,d^4z\;
    h_{\mu\nu}(x)\,h_{\alpha\beta}(y)\,h_{\rho\sigma}(z)\,
    \mathcal V_R^{\mu\nu,\alpha\beta,\rho\sigma}
    -\frac{\kappa}{2}\int d^4x\;
    h_{\mu\nu}(x)\,\mathcal T_{\rm pp}^{\mu\nu}(x),
    \label{eq:memory_eff_action}
\end{align}
with quadratic and cubic kernels
$\mathcal K_R=D_R^{-1}+(\kappa/2)^2\,\Pi_R$, with $\Pi_R$ the retarded
version of Eq.~\eqref{eq:Pi_def} \emph{including the seagull}, and
$\mathcal V_R=V_{{\rm EH},R}^{(3)}+\delta V_R$,
$\delta V_R\equiv(\kappa/2)^3\,\Gamma_{TTT,R}$.
Varying with respect to $h_{\mu\nu}$ gives the equation of motion
\begin{equation}
    \int d^4y\;
    \mathcal K_R^{\mu\nu}{}_{\alpha\beta}(x,y)\,
    h^{\alpha\beta}(y)
    =
    \frac{\kappa}{2}\,\mathcal T_{\rm pp}^{\mu\nu}(x)
    +
    J_{\rm EH}^{\mu\nu}[h,h](x)
    +
    J_{\delta V}^{\mu\nu}[h,h](x),
    \label{eq:memory_eom}
\end{equation}
with the nonlinear currents
\begin{equation}
    J_X^{\mu\nu}[a,b](x)
    \equiv
    -\frac{1}{2}
    \int d^4y\,d^4z\;
    V_{X,R}^{\mu\nu}{}_{\alpha\beta,\rho\sigma}(x,y,z)\,
    a^{\alpha\beta}(y)\,
    b^{\rho\sigma}(z),
    \qquad
    X={\rm EH},\delta V.
\end{equation}
Decompose the field into a hard radiative part $\bar h_{\mu\nu}$
($\omega\sim v/r$) and a soft memory part $\Delta h_{\mu\nu}$ ($\omega\to0$),
$h_{\mu\nu}=\bar h_{\mu\nu}+\Delta h_{\mu\nu}$, and write the retarded convolution
$(G_R * J)_{\mu\nu}(x)\equiv\int d^4y\,G^R_{\mu\nu;\alpha\beta}(x,y)\,
J^{\alpha\beta}(y)$. The soft memory field then obeys
\begin{equation}
    \Delta h_{\mu\nu}
    =
    \bigl(G_R * J_{\rm EH}[\bar h,\bar h]\bigr)_{\mu\nu}
    +
    \bigl(G_R * J_{\delta V}[\bar h,\bar h]\bigr)_{\mu\nu}.
    \label{eq:memory_master}
\end{equation}
To first order in the medium expansion,
$\mathcal T_{\rm pp}=\mathcal T_0+\delta\mathcal T$ and
$G_R=D_R+\delta G_R$ with
$\delta G_R=-D_R(\kappa/2)^2\Pi_R D_R+\mathcal O(\epsilon_{\rm env}^2)$,
so the hard radiation field splits into a vacuum part and the source- and
propagator-induced corrections,
\begin{equation}
    \bar h=\bar h_0+\delta\bar h_F+\delta\bar h_G,
    \qquad
    \bar h_0=\tfrac{\kappa}{2}D_R * \mathcal T_0,
    \quad
    \delta\bar h_F=\tfrac{\kappa}{2}D_R * \delta\mathcal T,
    \quad
    \delta\bar h_G=\tfrac{\kappa}{2}\delta G_R * \mathcal T_0,
    \label{eq:memory_hard_split}
\end{equation}
feeding the four-piece decomposition~\eqref{eq:memory_split} and the order
counting of Sec.~\ref{sec:app_memory}.

\section{Benchmarks: From Jeans Physics to Wave Dark Matter}
\label{app:benchmarks}

A framework that reorganizes known physics must return that physics on
demand---and should then produce new results with the same effort. This
appendix does both. Resumming the medium self-energy on the Newtonian
line reproduces the linearized self-gravitating (Jeans) dynamics of each
medium, with the gravitational sector normalized once and for all by
Newton's law. Alongside each benchmark, the same dressed propagator
yields closed-form wave-dark-matter observables, each either matching
the literature or, where indicated, new.

\paragraph{Static screening.}
The $00$--$00$ graviton carries the de~Donder weight $P_{0000}=\tfrac12$,
so each bulk-$\Pi$ insertion on the Newtonian line contributes the
relative factor $k_J^2/|\mathbf k|^2$ of
Eq.~\eqref{eq:newton_pi_relative}, derived on the Newton exchange in
Sec.~\ref{sec:Newton_Pi}. Resumming the geometric series,
\begin{equation}
    \mathcal A(\mathbf k)
    =\frac{\mathcal A_{\rm Newton}}{1-k_J^2/|\mathbf k|^2}
    =-\frac{4\pi G_N M_1 M_2}{|\mathbf k|^2-k_J^2},
\end{equation}
which is the Fourier transform of the modified Poisson equation
\begin{equation}
    \bigl(\nabla^2+k_J^2\bigr)\Phi(\mathbf x)=4\pi G_N\,\rho_{\rm ext}(\mathbf x),
    \qquad
    k_J^2=\frac{4\pi G_N(\epsilon+p)}{c_s^2},
\end{equation}
the standard linearized response of a
self-gravitating medium, with its pole at
the Jeans wavenumber~\cite{BinneyTremaine}. The de~Donder weight is exactly
what makes the resummed pole land at $k_J^2$ rather than $2k_J^2$, and
$k_J$ is an \emph{output} fixed by the medium, not a free infrared
parameter. The real pole at $|\mathbf k|=k_J$ reflects the Jeans
instability of the homogeneous background itself (the Jeans swindle);
only the static Green's function at and below $k_J$ therefore requires a
choice of global completion---a cosmological background, bounding
pressure, or background profile---beyond the local EFT.

\paragraph{Static screening in wave dark matter.}
For a coherent scalar the screening wavenumber is itself momentum
dependent: $c_s^2(\mathbf k)=\mathbf k^2/4m^2$ gives
$k_J^2(\mathbf k)=16\pi G_N\rho\,m^2/|\mathbf k|^2$. Dressing the static
exchange with this response turns the screened potential just derived into
\begin{equation}
    \Phi(\mathbf k)
    =-\frac{4\pi G_N M\,|\mathbf k|^2}{|\mathbf k|^4-k_{\rm dB}^4},
    \qquad
    k_{\rm dB}\equiv(16\pi G_N\rho\,m^2)^{1/4},
    \label{eq:waveDM_potential}
\end{equation}
with $k_{\rm dB}$ the de~Broglie (quantum-Jeans) scale. The Newtonian
$1/|\mathbf k|^2$ is recovered for $|\mathbf k|\gg k_{\rm dB}$, while the
poles of the quartic denominator produce a Yukawa-screened and an
oscillatory component. In position space, with the symmetric
(standing-wave) prescription for the real poles,
\begin{equation}
    \Phi(r)=-\frac{G_N M}{2r}\Bigl[\cos(k_{\rm dB}\,r)+e^{-k_{\rm dB}\,r}\Bigr],
    \label{eq:waveDM_potential_position}
\end{equation}
which recovers $-G_N M/r$ as $r\to0$ and leaves, beyond the de~Broglie
wavelength, the slowly decaying oscillatory tail
$-(G_N M/2r)\cos(k_{\rm dB}\,r)$---the imprint of the granular, wave-like
medium on the two-body force. Wave-dark-matter potentials and soliton
cores are well studied~\cite{Hu:2000ke,Hui:2016ltb}; what is new is that
this closed form follows directly from the dressed graviton propagator,
with no separate solution of the Schr\"odinger--Poisson system (the
infrared caveat above applies at and below $k_{\rm dB}$).

\paragraph{Dynamical pole.}
Restoring the frequency dependence, Eq.~\eqref{eq:Zs_def} with
$t_s^{00}=1$ gives the longitudinal self-energy
\begin{equation}
    \Pi^{0000}(\omega,\mathbf k)
    =-\frac{(\epsilon+p)\,|\mathbf k|^2}{\omega^2-c_s^2|\mathbf k|^2},
\end{equation}
and the gravitational dressing shifts the sound pole by the same $k_J^2$: the
dressed graviton in the scalar (longitudinal) sector---distinct from the
transverse-traceless mode, which is undressed at this order---propagates with
the relativistic Jeans--sound dispersion
\begin{equation}
    \boxed{\;
    \omega^2=c_s^2|\mathbf k|^2-4\pi G_N(\epsilon+p)\,,
    \;}
\end{equation}
where the inertial enthalpy $\epsilon+p$ is read off from
$\Pi^{0000}_{\rm static}$; for non-relativistic matter
($p\ll\epsilon\simeq\rho$) this is the textbook Jeans relation
$\omega^2=c_s^2k^2-4\pi G_N\rho$~\cite{BinneyTremaine}. The acoustic branch
and the gravitational instability emerge with their relative coefficient
\emph{fixed}, not fitted.

\paragraph{The three media.}
Specializing the single object $c_s^2$ then reproduces three independently
known results with no free parameter: a constant $c_s$ gives the classical
Jeans wavenumber $k_J^2=4\pi G_N\rho/c_s^2$; the scalar-field
$c_s^2(\mathbf k)=\mathbf k^2/4m^2$ [Eq.~\eqref{eq:cs_scalar}] turns
Eq.~\eqref{eq:jeans_dispersion} into
$\omega^2=\mathbf k^4/4m^2-4\pi G_N\rho$ with the self-consistent
$k_J=(16\pi G_N\rho\,m^2)^{1/4}$, the Hu--Barkana--Gruzinov quantum-Jeans
result~\cite{Hu:2000ke,Hui:2016ltb}; and the Landau response
[Eq.~\eqref{eq:Pi_collisionless}] gives the static gravitational dielectric
$\varepsilon(0,\mathbf k)=1-k_J^2/|\mathbf k|^2$ with
$k_J^2=4\pi G_N\rho/\sigma^2$, the kinetic Jeans
criterion~\cite{BinneyTremaine}. That all three fall out of the same
dressed propagator, each by a single replacement of the correlator, is
the central consistency check of the construction. Two further
wave-dark-matter observables close the circle.

\paragraph{Oscillating potential of a scalar condensate.}
The oscillating background pressure $p(t)=-\rho\cos(2mt)$
[Sec.~\ref{sec:media_catalogue}] feeds directly into the linearized
Einstein equations. In the conformal-Newtonian gauge
$ds^2=(1+2\Phi)dt^2-(1-2\Psi)d\mathbf x^2$, the absence of anisotropic
stress gives $\Phi=\Psi$, the constant $T^{00}$ forces $\delta\rho=0$, and
the trace equation reduces, for a homogeneous source, to
$\ddot\Phi=4\pi G_N\,p(t)$, so that
\begin{equation}
    \boxed{\;
    \Phi(t)=\Psi(t)=\frac{\pi G_N\rho}{m^2}\,\cos(2mt)\,.
    \;}
    \label{eq:KR_potential}
\end{equation}
This is precisely the Khmelnitsky--Rubakov
signal~\cite{Khmelnitsky:2013lxt}: a monochromatic $\sim\!10$--$100$\,ns
modulation of pulsar arrival times at
$f=m/\pi\approx4.8\,\mathrm{nHz}\,(m/10^{-23}\,\mathrm{eV})$---in the
present language, simply the metric sourced by the $2m$ harmonic of the
medium stress tensor.

\paragraph{Gravitational-wave dispersion through wave dark matter.}
The propagation counterpart is governed by the transverse-traceless graviton
self-energy. For a homogeneous condensate, moving or at rest, the coherent
part leaves the TT shell at $k^2=0$ at first order: on TT legs $g^{00}$ is
untouched and only the volume factor contributes, giving the same
$\Lambda$-type $-2p$ as the perfect fluid and no shear piece. The shift comes
from the kinetic part,
$\omega^2=|\mathbf k|^2+\tfrac{32\pi G_N}{3}\mathcal E$, with $\mathcal E$ the
kinetic energy density, the cold-dark-matter dispersion of Flauger and
Weinberg~\cite{Flauger:2017gtb}. The induced fractional speed shift is
negligible at gravitational-wave frequencies.

\bibliographystyle{JHEP}
\bibliography{references}

\end{document}